\documentclass[a4paper,11pt]{article}
\pdfoutput=1 
\usepackage{jcappub} 

\usepackage[T1]{fontenc} 

\usepackage{graphicx}
\usepackage{amsmath,amssymb,amsthm}
\usepackage{latexsym,graphicx,color,subfigure,slashed,mathrsfs,verbatim,bbm,wasysym,
  pstricks,epsfig,url, hyperref} 
\usepackage{epstopdf}
\usepackage{nicefrac}
\newcommand{\cO}{{\cal O}}

\newcommand{\beq}{\begin{eqnarray}}
\newcommand{\eeq}{\end{eqnarray}}

\newcommand{\cL}{{\cal L}}

\newcommand{\hL}{{\hat L}}
\newcommand{\htau}{{\hat \tau}}
\newcommand{\hQ}{{\hat Q}}
\newcommand{\hb}{{\hat b}}

\newcommand{\hnu}{{\hat \nu}}

\newcommand{\z}{\mathbb{Z}}

\title{The Fraternal WIMP Miracle}

\subheader{CERN-PH-TH-2015-125}

\author[a]{Nathaniel Craig}
\author[b,c]{and Andrey Katz}

\affiliation[a]{Department of Physics, University of California, Santa Barbara, CA 93106, USA}
\affiliation[b]{Theory Division, CERN, CH-1211 Geneva 23, Switzerland}
\affiliation[c]{Universit\'e de Gen\`eve, Department of Theoretical Physics and Center for Astroparticle Physics (CAP), 
24 quai E. Ansermet, CH-1211, Geneva 4, Switzerland}

\abstract{We identify and analyze thermal dark matter candidates in
  the fraternal twin Higgs model and its generalizations. The relic
  abundance of fraternal twin dark matter is set by twin weak
  interactions, with a scale tightly tied to the weak scale of the
  Standard Model by naturalness considerations. As such, the dark
  matter candidates benefit from a ``fraternal WIMP miracle,''
  reproducing the observed dark matter abundance for dark matter
  masses between 50 and 150~GeV. However, the couplings dominantly
  responsible for dark matter annihilation do not lead to interactions
  with the visible sector. The direct detection rate is instead set
  via fermionic Higgs portal interactions, which are likewise
  constrained by naturalness considerations but parametrically weaker
  than those leading to dark matter annihilation. The predicted direct
  detection cross section is close to current LUX bounds and presents
  an opportunity for the next generation of direct detection
  experiments.}

\begin{document}
\maketitle
\flushbottom

\section{Introduction}
\label{sec:intro}

Weakly Interacting Massive Particles (WIMPs) have long been a leading candidate to explain the enigma of dark matter (DM). The appeal of WIMP dark matter is due in part to the suggestive coincidence between the thermal abundance of WIMPs and the observed dark matter density, known as the ``WIMP miracle''. An important virtue of the WIMP miracle is that it connects the dark matter problem to the electroweak scale, where we expect to find new physics related to the hierarchy problem of the Standard Model (SM). Indeed, attempts to solve the hierarchy problem often lead to natural WIMP dark matter candidates.

However, in recent years the WIMP paradigm has come under pressure on two fronts. To date, direct searches for new physics at the Large Hadron Collider (LHC) have yet to yield evidence for new physics beyond the Standard Model. This calls into question the naturalness of the electroweak scale and naturalness-based motivation for new physics. In parallel, decisive exclusions from DM direct detection experiments such as LUX \cite{Akerib:2013tjd}, XENON100 \cite{Aprile:2012nq}, and CDMS \cite{Agnese:2013jaa} already go well beyond the scattering cross sections one might naively expect for a vanilla WIMP scenario. Although there are still several important 
loopholes in this argument (and substantial room for WIMPs with spin-dependent interactions), the parameter space for WIMP dark matter is increasingly constrained.

But perhaps this convergence of constraints provides a suggestive hint as to the nature of dark matter, opening the door to natural explanations of the weak scale that furnish a dark matter candidate without presenting conventional signatures at the LHC or direct detection experiments. In this respect scenarios of ``neutral naturalness'' such as the Twin Higgs~\cite{Chacko:2005pe} and Folded 
Supersymmetry~\cite{Burdman:2006tz} are promising alternatives. Such theories successfully improve the naturalness of the electroweak scale in spite of LHC bounds~\cite{Burdman:2014zta}. The key feature is that these models generically involve no new light colored states at the weak scale. Rather, the top partners in such theories are charged under a color group that is \emph{different} from the SM color, avoiding abundant LHC production of partner states and ubiquitous spectacular signatures. 

While neutral naturalness scenarios do not universally predict stable particles at the weak scale, they feature another nontrivial prediction, 
namely hidden valleys at scales 
of order $\cO(1-10)$~GeV. These hidden valleys~\cite{Strassler:2006im,Strassler:2006qa,Han:2007ae,Juknevich:2009ji} 
emerge in neutral naturalness models because the top partners must be charged under a non-SM 
$SU(3)$ color group. Typically, this $SU(3)$ group has a gauge coupling not very different from QCD, with modest amounts of fermionic matter charged under it. As such, this group typically confines at the scale of $1-10$~GeV, giving rise to a diverse spectrum of hadronic bound states. In addition to QCD-like interactions, natural twin Higgs models also possess twin analogs of the weak force to act as partner states for SM electroweak gauge bosons. Both the coupling and the breaking scale of this twin weak force are tightly connected to their SM counterparts. This provides a natural avenue for generalizing the WIMP miracle to dark matter candidates that obtain their thermal abundance not through weak interactions of the SM, but rather through weak interactions of hidden sectors related to the Standard Model by the twin Higgs mechanism. While the relic abundance of the dark matter candidate is set by twin weak annihilations, its direct detection rate is set by Higgs portal interactions connecting the twin sector to the Standard Model.

In this paper we study the generalization of the WIMP miracle in the context of the "Fraternal Twin Higgs"~\cite{Craig:2015pha}, the twin analogon of the "Natural SUSY" framework in supersymmetry~\cite{Dimopoulos:1995mi,Cohen:1996vb,Brust:2011tb,Papucci:2011wy}. 
Rather than requiring an exact mirror symmetry between the Standard Model and a twin sector as in the original twin Higgs model \cite{Chacko:2005pe}, the fraternal twin Higgs involves only the states and symmetries which are necessary for naturalness of the low-energy effective theory beneath a cutoff of order $\sim 5-10$~TeV. Such models respect a variety of global (and potentially gauge) symmetries that may serve to stabilize a dark matter candidate. In particular, we find that an accidental $U(1)$ symmetry of the fraternal twin sector automatically protects the lightest hidden lepton -- the twin tau -- making it a natural DM candidate.\footnote{Of course, twin Higgs models solve the {\it little} hierarchy problem and require a complete solution such as compositeness or supersymmetry to enter at the cutoff of the low-energy effective theory. These solutions may introduce additional dark matter candidates of their own, such as neutralinos in supersymmetric completions of the twin Higgs \cite{Chang:2006ra,Craig:2013fga}.}

We explore in detail the thermal relic prospects of the twin tau in fraternal twin Higgs models and their generalizations. We find that the relic abundance is largely determined by annihilations via the twin weak gauge bosons, providing a natural realization of a ``fraternal WIMP miracle'' in the twin sector. The preferred mass of the dark matter candidate can vary between 50 to 150~GeV, depending on the scale at which the accidental global symmetry protecting the Higgs boson is spontaneously broken. In contrast to dark matter annihilation, the direct detection rates are completely set by Higgs interactions. In this respect, the fraternal twin Higgs and related models are particular examples of a fermionic 
Higgs portal~\cite{Patt:2006fw, Kim:2006af} augmented by additional forces in the hidden sector. Intriguingly, the fraternal WIMP mechanism is already on the verge of being probed by direct detection 
experiments, and large portions of viable parameter space lie just beyond the current reach of the LUX experiment.

However, the fraternal WIMP miracle is far more general than the minimal twin Higgs mechanism. To this end, we study dark matter prospects in several natural extensions of the fraternal twin Higgs. Although there is no twin hypercharge gauge boson in the minimal fraternal twin Higgs, we first consider the consequences of weakly gauging the global $U(1)$ in the hidden sector. In this particular extension the twin taus are also allowed to annihilate into twin photons, favoring lighter masses (in the range of 1~to 10~GeV) for the thermal relic than in the minimal model. Such light dark matter is a challenge for existing direct detection experiments, although substantial progress has recently been made in this direction. Second, motivated by the orbifold UV completions of the twin 
Higgs~\cite{Craig:2014aea,Craig:2014roa}, we consider the dark matter in $\z_N$ extensions of the twin Higgs scenario. These $N$-Higgs models preserve the success of the fraternal WIMP miracle and necessarily lead to several coexisting dark matter species, possibly with different masses. The dominant relic abundance typically arises from the lightest twin lepton, but all DM populations can lead to signatures in direct detection experiments with different recoil spectra.

Last but not least, we emphasize that the fraternal twin higgs scenario and generalizations thereof are largely immune to the cosmological problems suffered by the original mirror twin Higgs scenario~\cite{Barbieri:2005ri}. By construction, the fraternal twin Higgs model has many fewer light degrees
of freedom than the mirror twin Higgs. If the Standard Model and twin sector decouple from one another early enough, the twin sector turns out to be much colder than the visible one. Thus the twin sector contributes less than one effective neutrino degree of freedom, well within current experimental bounds.

Our paper is structured as follows: In Sec.~\ref{sec:FraternalTwin} we briefly review the structure of the fraternal twin Higgs and describe in more detail its $\z_N$ generalizations. We also review the existing experimental constraints on this scenario from non-SM Higgs decays and Higgs coupling deviations. In Sec.~\ref{sec:ThermalHistory} we very briefly review the thermal history of the universe in the fraternal twin Higgs scenario with an emphasis on its cosmological safety relative to the mirror twin Higgs. In Sec.~\ref{sec:DM} we calculate in detail the expected thermal relic abundance in the fraternal 
twin Higgs and its extensions. In Sec.~\ref{sec:detection} we determine the expected nuclear scattering cross section for twin dark matter and compare it to current direct detection bounds, finding that this scenario is not meaningfully constrained by current experiments but largely within reach of future measurements. We conclude and discuss future directions in Sec.~\ref{sec:conclusions}.

\section{The Fraternal Twin Higgs and Its Cousins }
\label{sec:FraternalTwin}

We begin by reviewing the minimal ingredients of a fraternal twin Higgs model, following \cite{Craig:2015pha}. We also discuss simple generalizations, along the lines of orbifold Higgs models \cite{Craig:2014aea, Craig:2014roa}. In each case, hidden sectors related to the Standard Model by approximate discrete symmetries serve to improve naturalness of the weak scale, providing a wide range of potential dark matter candidates motivated by naturalness.

\subsection{The minimal fraternal twin Higgs}

In the original Twin Higgs model \cite{Chacko:2005pe}, the Standard Model is related to a mirror copy by an exact $\mathbb{Z}_2$ symmetry. Given this symmetry, the quadratic potential of the SM Higgs doublet and its mirror counterpart enjoys an enhanced $SU(4)$ symmetry.\footnote{In fact, in a perturbative model this $SU(4)$ symmetry is accidentally enhanced to a full $O(8)$ symmetry \cite{Chacko:2005un}, which is necessary to forbid certain dangerous operators from spoiling the twin mechanism \cite{Barbieri:2015lqa, Low:2015nqa}; we will not make meaningful use of this distinction here.} Spontaneous breaking of this accidental $SU(4)$ leads to a goldstone mode that may be identified with the SM-like Higgs boson, which then enjoys protection from $\mathbb{Z}_2$-symmetric threshold corrections at higher scales. 

However, not all features dictated by an exact $\mathbb{Z}_2$ symmetry are required for naturalness of the weak scale. The smallness of threshold corrections involving light generation yukawas or the hypercharge gauge coupling suggests that the related degrees of freedom may be decoupled in the mirror sector without rendering the Higgs unnatural. The appearance of incomplete multiplets of a symmetry in the low-energy effective theory is familiar from orbifolds and related constructions \cite{Dixon:1985jw}. This is not merely a choice of convenience designed to evade direct search limits, but a consequence of ancillary considerations. Much as the spectrum of ``Natural SUSY'' was originally motivated by flavor considerations~\cite{Dimopoulos:1995mi,Cohen:1996vb}, the spectrum of a so-called ``fraternal'' twin Higgs is in part motivated by cosmological considerations -- namely, the elimination of additional light degrees of freedom that might create tension with the observed bound on the effective number of neutrino species.

The necessary ingredients required for a natural fraternal twin Higgs were enumerated from the perspective of the low-energy effective theory in \cite{Craig:2015pha}, and we will merely summarize the conclusions here. The Standard Model is related to a hidden sector via an approximate $\mathbb{Z}_2$ symmetry that relates states with $\mathcal{O}(1)$ interactions. In addition to the field content of the Standard Model with a single complex scalar Higgs doublet $A$, a viable fraternal twin model entails a hidden sector containing
\begin{enumerate}
\item A twin Higgs doublet $B$, entirely neutral under the Standard Model. The $A$ and $B$ Higgs multiplets possess an approximately $SU(4)$-symmetric potential of the form $\lambda (|A|^2 + |B|^2)^2$, which itself enjoys an accidental $O(8)$ symmetry.
\item Twin tops and a twin top Yukawa that is numerically close to the SM top Yukawa -- within $\sim \pm 1\%$ for a cutoff of $\Lambda \sim 5-10$ TeV.
\item A gauged twin $SU(2)$ weak symmetry with twin weak $W$ and $Z$ bosons and couplings $\hat g_2(\Lambda) \approx g_2(\Lambda)$.
\item A gauged twin $SU(3)$ symmetry with twin gluons and $\hat g_3(\Lambda) \approx g_3(\Lambda)$. Crucially, this gauge group is asymptotically free and confines at a scale $\hat{\Lambda}$.
\item Twin bottoms and twin taus, necessary for anomaly cancellation; the masses of these states are essentially free parameters so long as they do not introduce tension with bounds on the non-SM Higgs width and remain much lighter than the twin top to avoid introducing new large threshold corrections.
\item A twin neutrino from the twin tau doublet, which may have a Majorana mass -- again a free parameter, as long as it is sufficiently light.
\end{enumerate}

Provided the approximate $\mathbb{Z}_2$ symmetry of these states and $\mathbb{Z}_2$-symmetric contributions above the cutoff of the low-energy effective theory, one-loop radiative corrections to the $A$ and $B$ multiplets assemble themselves into an approximate $SU(4)$ invariant $\propto |A|^2 + |B|^2$. Given an approximately $SU(4)$-symmetric potential for the Higgs multiplets of the form 
\begin{equation}
V \supset \lambda (|A|^2 + |B|^2)^2,
\end{equation}
nonzero vacuum expectation values (vevs) for $A$ and $B$ of the form $|\langle A \rangle|^2 + |\langle B \rangle|^2 = \frac{f^2}{2}$ lead to spontaneous breaking of the $SU(4)$ symmetry with order parameter $f$ (as well as breaking of electroweak symmetry with order parameter $v$). Six of the seven resulting (pseudo-)goldstone bosons are eaten by the Standard Model and twin weak gauge groups, leaving behind a real pseudo-goldstone boson that may be identified with the observed SM-like Higgs. As a goldstone of spontaneous $SU(4)$ breaking, its mass is protected against $\mathbb{Z}_2$-symmetric threshold corrections at the cutoff scale $\Lambda$. 

In contrast with models involving continuous symmetries at the weak scale such as supersymmetry or composite Higgs models, in this case the light partner states -- the fermions and gauge bosons outlined above -- are not charged under the Standard Model gauge interactions. As such, they are not copiously produced at the LHC. The primary bounds arise at present from $\mathcal{O}(v/f)$ Higgs coupling deviations, since the SM-like Higgs is a pseudo-goldstone boson with $\mathcal{O}(v/f)$ mixing with the twin sector Higgs. 

From the perspective of dark matter, there is a wide range of potential dark matter candidates, ranging from the lightest fermionic states in the twin sector to potential bound states of twin QCD confinement. The crucial point from the perspective of the WIMP miracle is that these states generically possess ``fraternally weak-scale'' interactions, set by the scale $\hat{g}_2 f$ rather than $g_2 v$. However, from the perspective of direct detection, their interactions with Standard Model states proceed through a Higgs portal coupling with parametrically different scaling. This Higgs portal coupling can be thought of as arising from mixing between the $A$ and $B$ sector Higgs doublets, or equivalently by integrating out the radial mode of spontaneous $SU(4)$ symmetry breaking. In particular, the twin tau $\htau$ couples to the SM-like Higgs via an interaction of the form
\begin{equation} 
\label{eq:yukawa}
\mathcal{L} \supset  \frac{m_{\hat{\tau}}}{f} \frac{v}{f} h \hat \tau \hat{\bar{\tau}} + {\rm h.c.},
\end{equation}
keeping in mind that $m_{\hat \tau}$ is essentially a free parameter. This raises the prospect of dark matter candidates whose abundance is set by a fraternal analog of the WIMP miracle, yet whose nuclear scattering cross sections lie outside the reach of current direct detection experiments.

\begin{figure}[t]
\centering
\includegraphics[width=.5\textwidth]{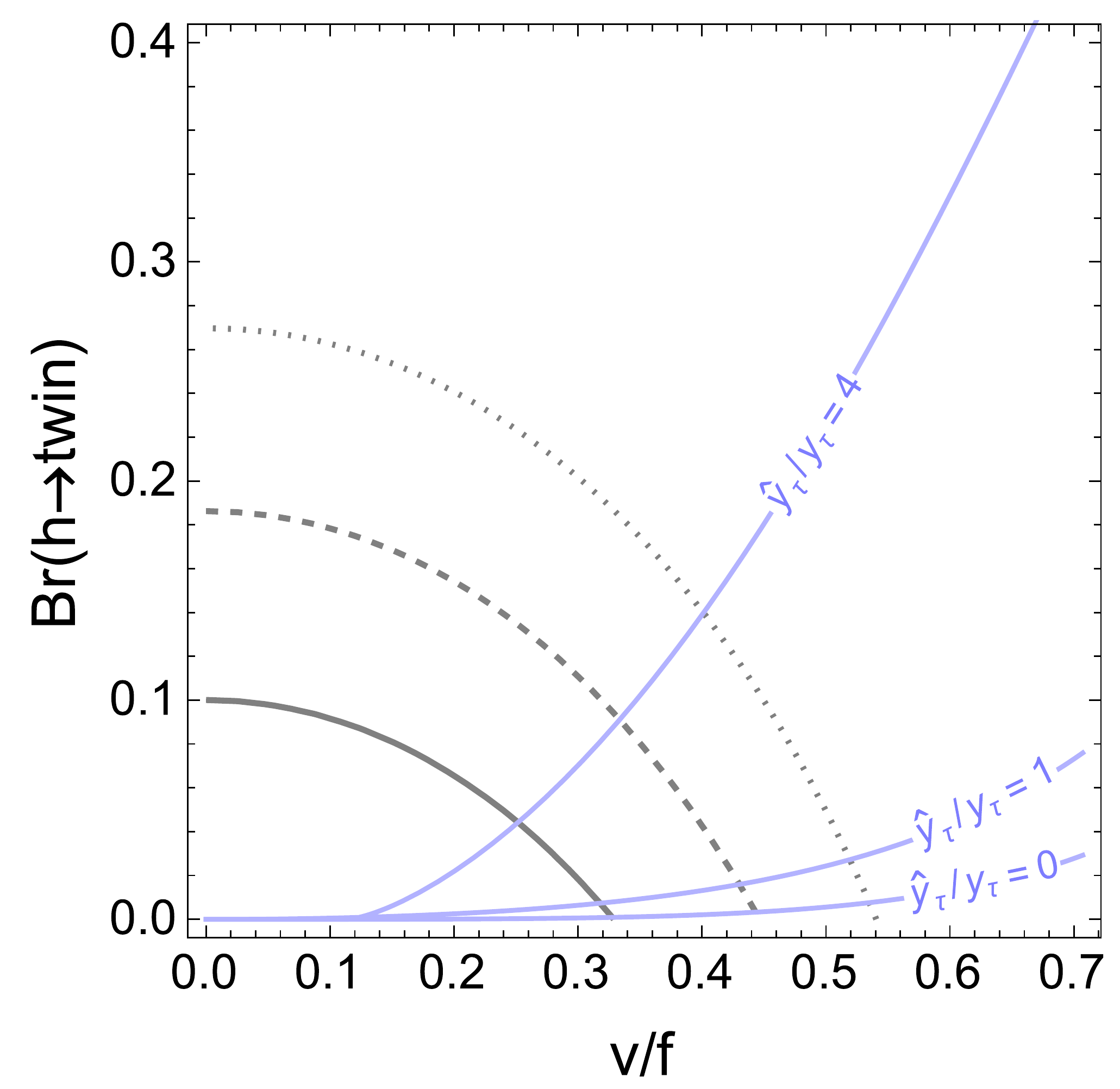}
\caption{Bounds on $v/f$ and the Higgs invisible branching ratio from a fit to current Higgs coupling measurements. The solid, dashed, and dotted black lines denote the 1-,2-, and 3-sigma bounds (defined as $\Delta \chi^2 = 2.30, 6.18,$ and $11.83$, respectively). The blue lines correspond to the contribution of Higgs decays into $\htau$ pairs as a function of $v/f$ for $\hat{y}_\tau = 0,1,4$. Here we neglect possible contributions from decays into twin bottoms, consistent with the freedom to have $\hat{y}_b \simeq 0$, but include irreducible contributions from decays into twin gluons, assuming twin glueballs are kinematically accessible. }
\label{fig:higgsfit}
\end{figure}

As we will see in subsequent sections, a natural dark matter candidate in fraternal twin Higgs models is the twin tau, whose stability is guaranteed by the gauged or global twin hypercharge symmetry. The mass of the twin tau is not necessarily fixed in fraternal models, since one-loop threshold corrections arising through the tau sector are negligible unless the twin tau yukawa is exceptionally large. However, as mentioned above, an experimental bound on the twin tau yukawa arises from constraints on the non-SM width of the Higgs boson. 

To quantify this constraint with an eye towards understanding the implications for dark matter, we plot the allowed parameter space in the plane of $v/f$ and Higgs invisible width in Fig.~\ref{fig:higgsfit}. Our fit procedure is the same as in \cite{Craig:2015pha}, albeit with the inclusion of ATLAS direct search limits on invisible decays of the Higgs. We overlay on this the contribution of Higgs decays into $\htau$ pairs as a function of $v/f$ for $\hat{y}_\tau = 0,1,4$. In doing so, we neglect possible contributions to the non-SM Higgs width from decays into twin bottoms, since the twin bottom mass is also a relatively free parameter and may be entirely negligible. For $v/f = 1/3$, bounds on the non-SM Higgs width constrain $\hat{y}_\tau \lesssim 4$, corresponding to $m_{\hat \tau} \lesssim 21$ GeV. However, for smaller values of $v/f$, the twin tau mass increases relative to $v/f = 1/3$ for fixed values of  $\hat{y}_\tau$ (due to the higher scale of symmetry breaking in the twin sector), such that the partial width into twin taus rapidly becomes threshold suppressed. For even $v/f = 1/4$, threshold suppression of $h \to \htau \htau$ is sufficiently strong that current Higgs coupling constraints place no bound on $\hat{y}_\tau$. 

In summary, these theories consist of mirror twin top quarks with a Yukawa closely related (within about $\pm 1 \%$) to the Standard Model top Yukawa; twin $W$ and $Z$ bosons with gauge couplings fairly close (within about $\pm 10\%$) to the Standard Model weak gauge coupling; twin gluons with gauge coupling on the same order (within about $\pm 30\%$) as the QCD coupling; twin bottoms and taus with masses unconstrained by naturalness provided $m_{\hat b}, m_{\hat \tau} \lesssim m_t$, but experimentally constrained by Higgs coupling measurements; and a twin neutrino with relatively free mass. Of these states in the twin sector, the twin taus provide the most natural dark matter candidate. The scale of the twin sector is set by the scale of $SU(4)$ symmetry breaking $f$, which also controls deviations in Higgs couplings proportional to $v/f$. The completely natural case $v \sim f / \sqrt{2}$ is excluded by Higgs coupling measurements, which constrain $v / f \lesssim 1/3$. The most natural surviving parameter space has $v/f$ as large as possible consistent with these constraints, with the fine-tuning of the weak scale proportional to $f^2 / v^2$. In what follows we will largely focus on the region of parameter space with $v/f \sim 1/5 - 1/3$, which corresponds to fine-tuning on the order of 5$-$20\%.

\subsection{Generalizations of the fraternal twin Higgs}

The twin Higgs mechanism admits a wide variety of generalizations with more elaborate symmetry structures. These extensions are naturally organized as orbifolds of theories with continuous symmetries \cite{Craig:2014aea, Craig:2014roa}, leading to multiple hidden sectors coupled through the Higgs portal whose combined matter content stabilizes the Higgs potential. These include the simple $\mathbb{Z}_n$ generalizations of the $\mathbb{Z}_2$ symmetry present in the fraternal twin Higgs, as well as much more elaborate configurations arising from non-abelian orbifolds whose low-energy degrees of freedom are not related by simple discrete symmetries \cite{Craig:2014roa}. As in the case of the fraternal twin Higgs, these theories furnish natural dark matter candidates whose thermal relic abundances arise in an analogous matter, illustrating the generality of the fraternal WIMP miracle.

Although a wide variety of extensions of the fraternal twin Higgs are possible, for simplicity we focus on the fraternal triplet Higgs, which suffices to illustrate the general mechanism. While the details of this theory were sketched in \cite{Craig:2014roa}, we review them here with a particular eye towards implications for dark matter candidates. The theory consists of three sectors related by an approximate $\mathbb{Z}_3$ symmetry, in a straightforward generalization of the $\mathbb{Z}_2$ scenario discussed above. In addition to the Standard Model with Higgs doublet $A$, there are two additional hidden sectors with Higgs doublets $B$ and $C$, respectively. In direct analogy with the fraternal twin Higgs, each hidden sector also minimally contains its own twin tops \& bottoms, twin tau and tau neutrinos, twin weak bosons, and twin glue. The scalars $A,B,C$ share an approximately $SU(6)$-symmetric potential of the form
\begin{equation}
V \supset \lambda (|A|^2 + |B|^2 + |C|^2)^2
\end{equation}
When these Higgs multiplets obtain vacuum expectation values, the approximate $SU(6)$ global symmetry of the Higgs sector is spontaneously broken, resulting in several light pseudo-goldstone bosons. To obtain a light Standard Model-like pseudo-goldstone Higgs, we consider the case $\langle A \rangle \ll \langle B \rangle, \langle C \rangle$. In this limit the scalar spectrum consists of a heavy radial mode, a Standard Model-like PNGB Higgs, and a second PNGB Higgs coupled very weakly to Standard Model fermions.

To study the physics of the light degrees of freedom, we can integrate out the radial mode, working in the limit $\langle A \rangle \to 0$. In general, in this limit both $B$ and $C$ acquire vevs, with $|\langle B \rangle|^2 + |\langle C \rangle|^2 = \frac{1}{2} f^2$; the precise distribution of the vevs depends on the details of the Higgs potential. We can parameterize the range of possible vevs by an angle $\beta$, such that $\langle B \rangle = \sin(\beta) f / \sqrt{2}, \langle C \rangle = \cos(\beta) f/\sqrt{2}$. The radial mode of spontaneous $SU(6)$ symmetry breaking is the real scalar fluctuation around the total vev $f$.

To obtain the couplings of the radial mode, we rotate the neutral scalars from the gauge eigenbasis to the ``Higgs basis'',
\begin{equation}
\left(\begin{array}{c} R \\ G \end{array} \right) = \left( \begin{array}{cc} s_\beta & c_\beta \\ -c_\beta & s_\beta \end{array} \right) \left( \begin{array}{c} B^0 \\  C^0 \end{array} \right)
\end{equation}
where $R$ is the real scalar radial mode and $G$ is the additional pseudo-goldstone mode (aside from the SM-like pseudo-goldstone $h$). The Higgs basis coincides with the mass eigenbasis in the limit $\langle A \rangle \to 0$; nonzero $\langle A \rangle$ leads to small $\mathcal{O}(v/f)$ corrections to this picture. We can obtain the leading couplings of the SM-like Higgs to fermions in the fraternal hidden sectors upon integrating out the radial mode by using the relation
\begin{equation}
|A|^2 + |B|^2 + |C|^2 = \frac{1}{2} f^2 \to R = f - \frac{h^2}{2f} - \frac{G^2}{2f} + \dots
\end{equation}

It's straightforward to work out the implications for naturalness in this scenario. Before spontaneous symmetry breaking, the yukawa couplings in the $B$ and $C$ sectors are of the form
\begin{equation}
\mathcal{L} \supset - y_t  B \hat Q_B \hat{\bar{u}}_B - y_t C \hat Q_C \hat{\bar{u}}_C + {\rm h.c.}
\end{equation}
with analogous yukawas for the fraternal leptons and bottom-type quarks. After spontaneous symmetry breaking and integrating out the radial mode, the SM-like Higgs $h$ inherits couplings to fraternal fermions of the form
\begin{equation} \label{eq:z3eff}
\mathcal{L} \supset  \frac{y_t s_\beta}{2 \sqrt{2} f} h^2 \hat u_B \hat{\bar{u}}_B + \frac{y_t c_\beta}{2 \sqrt{2} f} h^2 \hat u_C \hat{\bar{u}}_C + \dots =  \frac{y_t^2 s_\beta^2}{4 m_{\hat{t}_B}} h^2 \hat u_B \hat{\bar{u}}_B  + \frac{y_t^2 c_\beta^2}{4  m_{\hat{t}_C}} h^2 \hat u_C \hat{\bar{u}}_C + \dots
\end{equation}
where $m_{\hat{t}_B} = y_t s_\beta f / \sqrt{2}, m_{\hat{t}_C} = y_t c_\beta f /\sqrt{2}$ are the fermion masses resulting from spontaneous symmetry breaking.

We can now see clearly the sense in which these hidden sectors improve the naturalness of the Higgs boson in the low-energy theory. Working in an effective theory with hard momentum cutoff $\Lambda$, the two couplings in Eq.~\ref{eq:z3eff} give a radiative correction to the Higgs mass that is quadratically divergent when we use a single mass insertion to close the loop, while higher mass insertions give the finite threshold correction to the Higgs mass. The combination of the two diagrams gives a correction to the Higgs mass
\begin{equation}
\delta m_H^2 = \frac{12}{16 \pi^2} \frac{y_t^2}{2} (s_\beta^2 + c_\beta^2) \Lambda^2 = +  \frac{6y_t^2}{16 \pi^2} \Lambda^2
\end{equation}
which is precisely the coefficient required to cancel the quadratic divergence from the top loop. The top quarks of the two hidden sectors combine to act as fermionic top partners in precisely the manner required for global symmetry protection of the Higgs.

Such extensions of the minimal fraternal twin Higgs are particularly interesting from the perspective of dark matter. In principle, both hidden sectors in the $\mathbb{Z}_3$ theory possess viable dark matter candidates in the form of the fraternal taus $\hat \tau_B, \hat \tau_C$ (although more elaborate possibilities such as asymmetric dark matter are also possible). The mass of each fraternal tau is a free parameter, while the Higgs portal coupling in terms of the physical masses is
\begin{equation} 
\mathcal{L} \supset  \frac{m_{\hat{\tau}_B}}{f} \frac{v}{f} h \hat \tau_B \hat{\bar{\tau}}_B + \frac{m_{\hat{\tau}_C}}{f} \frac{v}{f} h \hat \tau_C \hat{\bar{\tau}}_C + {\rm h.c.},
\end{equation}
a straightforward generalization of the fraternal twin Higgs. This attests to the breadth of the fraternal WIMP mechanism, in the sense that generalizations of the fraternal twin Higgs likewise possess parametrically WIMP-like interactions within each sector and Higgs portal couplings to the Standard Model.

\section{Thermal History of the Fraternal Twin Higgs}
\label{sec:ThermalHistory}

Having laid the groundwork for the fraternal twin Higgs and its generalizations, we now turn to the resulting cosmology. Before discussing thermal dark matter candidates in detail, we first sketch the broad outlines of the thermal history of the universe in a fraternal twin scenario. For simplicity, we assume the Standard Model and fraternal twin sector(s) are uniformly reheated (though much of the following discussion could be avoided entirely in theories with sufficiently low reheating temperatures). After reheating, the Standard Model and twin sectors are kept in thermal equilibrium via the Higgs portal quartic coupling $\lambda$. This is in contrast to mirror scenarios unrelated to naturalness, in which the Higgs portal quartic interaction is often taken to be infinitesimally small in order to avoid thermal equilibration between the SM and mirror sectors \cite{Ignatiev:2000yy}. Here the quartic must be $\mathcal{O}(1)$ in order for the pseudo-goldstone SM-like Higgs to be parametrically lighter than other states associated with the spontaneous breaking of approximate global symmetries. Consequently, interactions between light Standard Model states and light twin states are mediated by both the SM-like Higgs and the radial Higgs mode. 

As the universe cools, the Higgs portal interactions eventually decouple, while the light degrees of freedom in each sector remain in equilibrium among themselves due to interactions mediated by their respective gauge bosons. We can crudely determine the decoupling temperature between the SM and twin sectors by focusing on when the twin taus drop out of thermal equilibrium with the SM; this happens when
\begin{equation}
n_{\hat \tau} \langle \sigma v \rangle \simeq H
\end{equation}
which occurs around $T_D \sim 1-3$ GeV for $m_{\hat \tau} \sim 10-100$ GeV.\footnote{Here we have assumed there is no light twin photon, in keeping with the minimal fraternal twin Higgs scenario. If there is a light twin photon, it provides additional portals for thermal equilibrium between the SM and twin sectors via both kinetic mixing and Higgs portal interactions. As we will see, for the parameter space of interest these effects are typically unimportant.} For simplicity we also assume the twin sector gluons decouple at the same temperature, which is reasonable given that the confinement scale is in this range. If we had additional light states in thermal equilibrium in the twin sector, the twin tau neutrino would decouple from this twin thermal bath around 6 MeV, slightly sooner than in the SM because of the heavier twin gauge bosons. Note that we are assuming no appreciable mixing between the twin neutrino and Standard Model neutrinos, which might otherwise provide another avenue for equilibration between the neutrino species. At this point if the temperatures of the Standard Model and twin sectors remained comparable, the contribution of the twin neutrino to the effective number of neutrino species would be in conflict with current limits.

However, at this stage the Standard Model possesses far more light degrees of freedom than the fraternal twin sector, which lead to successive reheatings of the Standard Model sector as they decouple. As the universe cools further from $T_D \sim 1$ GeV down to below the scale of electron decoupling, various SM states come out of thermal equilibrium at or below the QCD phase transition. In the twin sector there is only the twin neutrino beneath the scale of the twin tau and twin glueballs. The entropy transfer from these SM decoupling phases reheats the Standard Model relative to the twin neutrinos, such that $T_{\hnu} \simeq (10.75/75)^{1/3} T_\nu$, where $g_* \simeq 75$ when the twin neutrinos decouple from the SM bath and $g_* = 10.75$ when the SM neutrinos decouple.

The strongest constraint on light degrees of freedom in the twin sector arises from the most recent combined CMB bound on light species, $N_{\rm eff} = 3.15 \pm 0.23$ \cite{Ade:2015xua}. Thanks to the preferential reheating of the Standard Model relative to the twin sector, the additional twin neutrino species should contribute $\Delta N_{\rm eff} \approx 0.075$, well within the allowed range. Thus we can conclude that, because the twin sector comes out of thermal equilibrium with the Standard Model early enough, the entropy deposition from decoupling of various light SM species results in much colder twin neutrinos, such that the bound from $N_{eff}$ is readily satisfied.

The stage is now set for studying the thermal history of the twin tau, which is dominantly set by the freezeout of its annihilation into twin neutrinos, as well as its Higgs portal annihilation into the Standard Model.

\section{Dark Matter Candidates and Thermal Relic Abundance}
\label{sec:DM}
\subsection{The basic picture}
Fraternal twin Higgs models provide fertile ground for dark matter candidates, insofar as they require additional hidden sectors neutral under the Standard Model with several possible gauge and global symmetries. If hypercharge is gauged in the twin sector, twin charge conservation provides a natural candidate for a dark matter stabilization symmetry. However, even if twin hypercharge is not gauged (or gauged and broken at a higher scale), the minimal fraternal twin Higgs model still possesses an accidental global $U(1)$ symmetry. The twin Higgs has $1/2$ charge under this symmetry, while the twin fermions have charges
\beq
\hL \sim -\frac{1}{2}, \ \ \ \htau^c \sim 1, \ \ \ \hQ \sim \frac{1}{6}, \ \ \ \hb^c \sim \frac{1}{3} 
\eeq 
respectively. When the twin electroweak phase transition occurs, both twin $SU(2)$ and the $U(1)$ symmetry are broken by the Higgs vev but, as in the SM, the generator 
$\hat Y + \hat T_3$ remains unbroken.  This is of course completely analogous to electromagnetism in the Standard Model, albeit at the level of a global symmetry in the twin sector.

Of course, if twin hypercharge is not gauged, then it is not plausible to expect that the global $U(1)$ symmetry remains an exact symmetry in the ultraviolet. Even if the field 
theory UV completion preserves the twin hypercharge, it should generically be broken by Planck-suppressed operators. Nonetheless, it is feasible that a discrete subgroup 
of the hypercharge remains unbroken and in the low-energy effective theory a discrete subgroup of  twin electrognetism remains exact. Therefore, even if 
we do not gauge twin hypercharge, we can expect that the lightest particle carrying the (global) twin electromagnetic charge is stable. This particle is a natural DM candidate. 

The case for dark matter stabilized by twin electromagnetic charge becomes even more compelling if the twin hypercharge global symmetry is weakly gauged. In this scenario, twin electromagnetism is an exact gauged symmetry and the lightest EM-charged particle must be stable. As explained 
in~\cite{Craig:2015pha} this gauging can radically change the collider signatures of the model, but as we will see here it has only a minor effect on the dark matter relic abundance and direct detection in a wide range of parameter space.  

\begin{figure}[t]
\centering
\includegraphics[width=.49\textwidth]{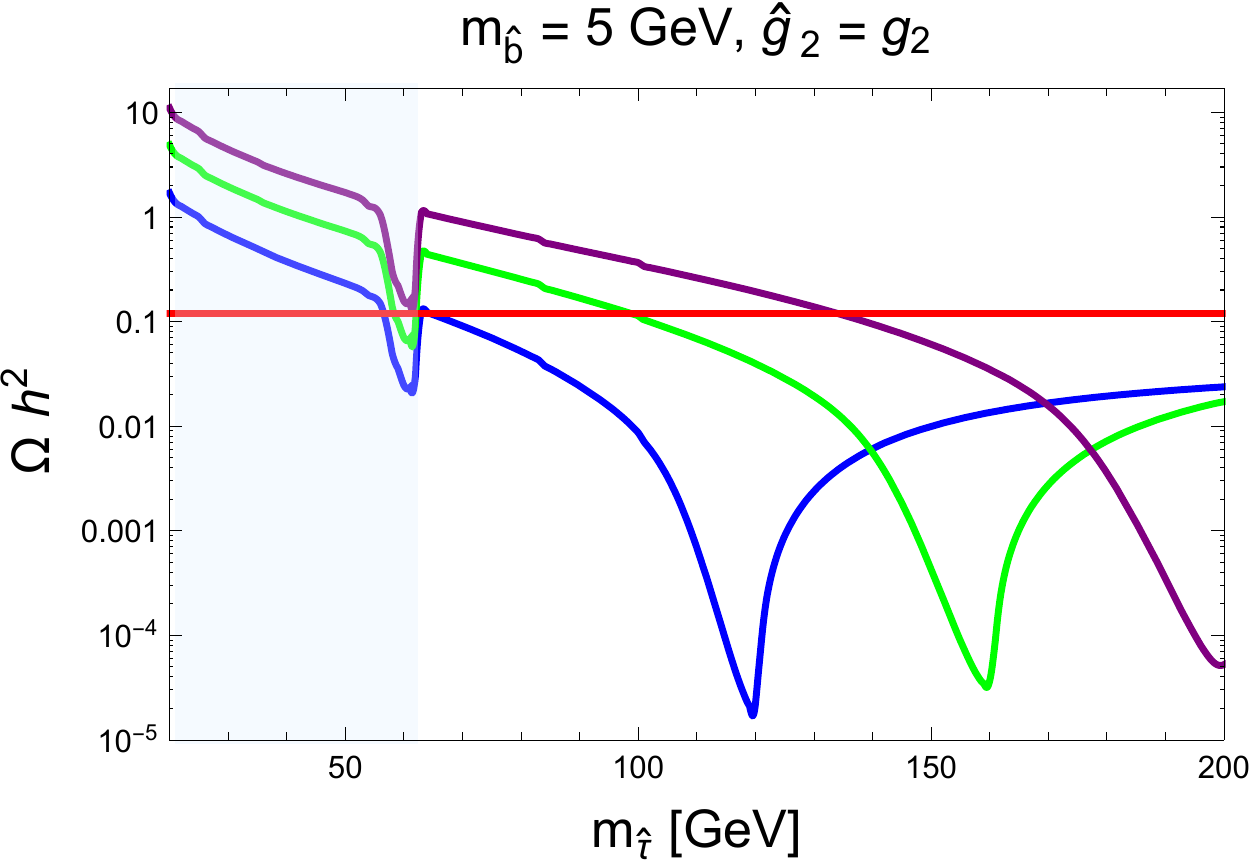}
\includegraphics[width=.49\textwidth]{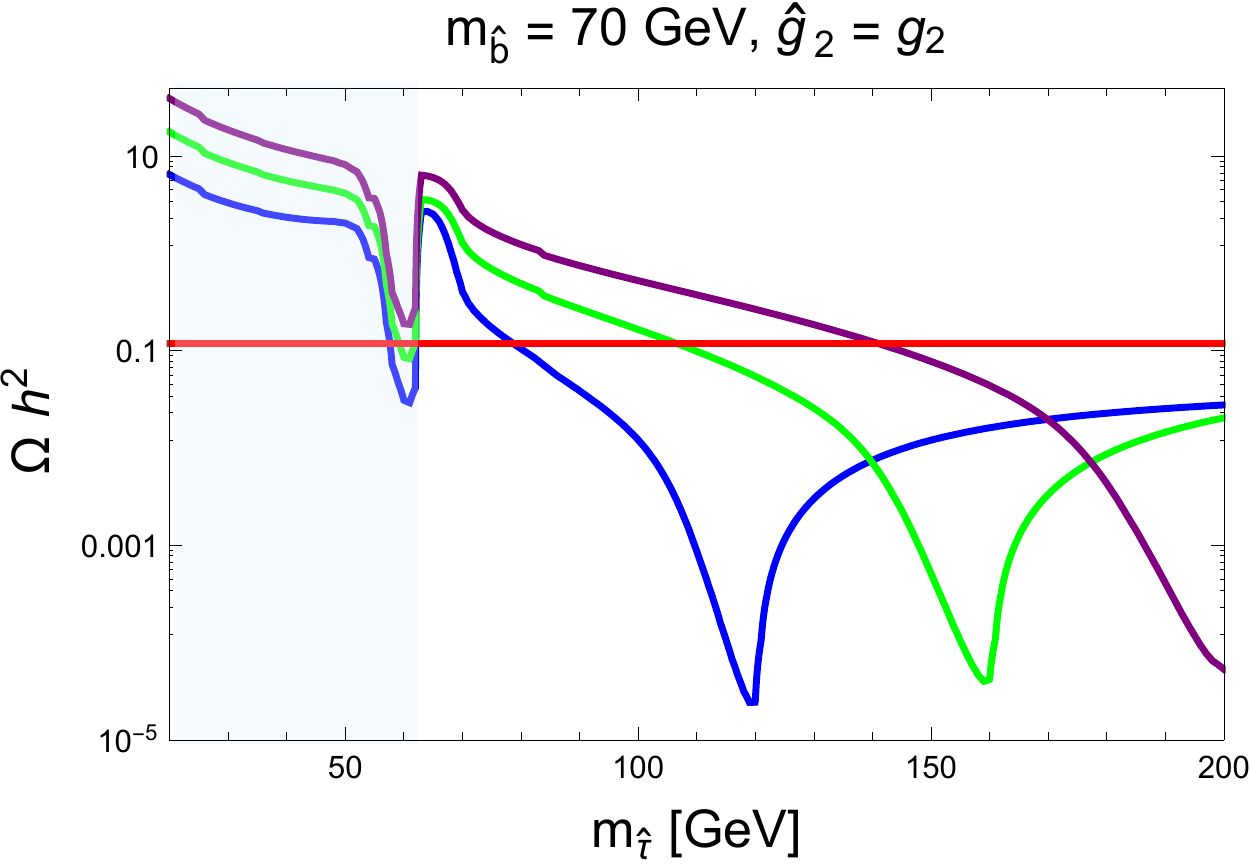}
\caption{Thermal relic abundance of the twin tau $\htau$ in a fraternal twin Higgs scenario without gauging twin hypercharge.
The blue, green  and purple line stand for $v/f = 1/3,\ 1/4,\ 1/5$ respectively.  The red line indicates the measured relic abundance $\Omega_{DM} h^2 = 0.1187.$ The light blue shading indicates the region excluded by Higgs coupling measurements for $v/f = 1/3$ only. Left: Relic abundance of $\htau$ assuming a relatively light twin bottom quark $\hat b$ and low scale of twin confinement. Right: Relic abundance of $\htau$ assuming a heavy twin bottom quark $\hat b$ and high scale of twin confinement. As we see the dependence on the exact mass spectrum of the twin particles is relatively weak. }

\label{fig:oh2noY}
\end{figure}

In the fraternal twin Higgs, the natural candidates for lightest fermion carrying (gauge or global) twin electromagnetic charge are the twin tau and twin b-quark. The masses of these fermions are in principle free parameters (so long as their yukawas are not large enough to pose a threat to naturalness at one loop), and  therefore either of them can be the lightest electrically charged particle.
However, due to the confinement of twin color, at low energy the b-quarks are not present as free particles, but rather a complicated spectrum of electromagnetically neutral twin quarkonia states. Therefore the likeliest DM candidate stabilized by twin electromagnetic charge in the hidden sector is a twin tau, and we will focus on this case in what follows.\footnote{Apart from (gauged or global) twin electromagnetism, the twin sector also possesses accidental twin lepton and baryon number symmetries. These symmetries are good accidental symmetries at the level of the light degrees of freedom, but more generally higher-dimensional operators connecting the Standard Model and twin sector lead to a single set of conserved baryon and lepton number symmetries. If twin baryon and lepton number are good symmetries, they may also serve to stabilize a dark matter candidate. This requires production of matter-antimatter asymmetry in the hidden sector, leading naturally to asymmetric
DM~\cite{Kaplan:2009ag}. We will not study this possibility here.}

If the twin hypercharge is not gauged, twin taus annihilate via twin weak gauge bosons into twin neutrinos. The differential cross section is 
\beq
\frac{d \sigma (\htau \htau \to \hnu \hnu)}{d \cos \theta}  = \frac{\hat{g}_2^4}{1024 \pi} \sqrt{\frac{s^2 m_\htau^2}{s - 4 m_\htau^2}} \left( 1 + c_\theta \sqrt{1 - \frac{4 m_\htau^2}{s}} \right)^2 \left| \frac{1}{s - m_{\hat{W}}^2} - \frac{2}{t - m_{\hat W}^2} \right|^2
\eeq 
where $t = m_\htau^2 - \frac{s}{2} \left(1 - c_\theta \sqrt{1 - \frac{4 m_\htau^2}{s} } \right)$, we approximate the resonant region with a Breit-Wigner distribution in the $s$-channel, and we have neglected the mass of the twin neutrinos. If the mass of the twin tau is much smaller
than the scale of twin weak bosons $gf/2$, the annihilation cross section takes the relatively simple form
\beq
\sigma(\htau \htau \to \hnu \hnu) \approx \frac{1}{4\pi} \sqrt{\frac{m_\htau^2}{s-4m_\htau^2}} G_F^2 \left(\frac{v}{f}\right)^4
\left( \frac{s}{4}\right) \left( 1+ \frac{1}{3} \left( 1-\frac{4m_\htau^2}{s}\right)\right)
\eeq

If the the twin b-quarks are lighter than twin taus and the decoupling temperature is higher 
that the twin QCD scale, there is an extra contribution to the twin tau annihilation rate due to annihilation into $\hb$ pairs, which then confine and eventually decay into the Standard Model:
\beq
\frac{d \sigma (\htau \htau \to \hb \hb) }{d \cos \theta} = \frac{3 \hat{g}_2^4}{1024 \pi} \sqrt{ \frac{s m_\htau^2(s - 4 m_\hb^2)}{s- 4 m_\htau^2}} \left( 1 + c_\theta \sqrt{1 - \frac{4 m_\hb^2}{s}} \sqrt{1 - \frac{4 m_\htau^2}{s} } \right)^2 \left| \frac{1}{s - m_{\hat{W}}^2} \right|^2
\eeq 
where we again approximate the resonant region with a Breit-Wigner distribution in the $s$-channel. Again when the mass of the twin tau is much smaller than the scale of the weak bosons, the cross section takes the simple form

\beq\label{eq:AnnBs}
\sigma(\htau \htau \to \hb \hb) \approx \frac{3}{4\pi} 
\sqrt{\frac{m_\htau^2}{s-4m_\htau^2}} G_F^2 \left(\frac{v}{f}\right)^4
\left( \frac{s}{4}\right) \left( 1+ \frac{1}{3} \left( 1-\frac{4m_\htau^2}{s}\right) \left( 1-\frac{4m_\hb^2}{s}\right)\right)
\eeq

On top of the annihilation of twin taus into the hidden sector as described above, the twin tau can also annihilate via Higgs resonances into SM states -- primarily
into $b \bar b$, but also $W^+W^-$, $ZZ$, and $hh$ if kinematically allowed. These annihilations  are suppressed by the SM bottom Yukawa 
(if annihilations into the gauge bosons are kinematically disallowed) and are also velocity suppressed~\cite{Fedderke:2014wda}.\footnote{Higgs-mediated processes may make significantly larger contributions to $\langle \sigma v\rangle$ if the dark sector is augmented with extra sources of CP violation and couplings of the DM candidate to the Higgs of the form $\sim \bar \chi (i\gamma^5 ) \chi |H|^2$ arise. This does not happen in the simplest fraternal twin Higgs scenario that we analyze here, but can happen in more complicated theories. If this coupling
is present, the theory might lose some of the features of the "fraternal WIMP miracle" and instead proceed along the lines of the ferminic Higgs portal
with non-standard Yukawa couplings~\cite{Bishara:2015cha}. }
Therefore, in most portions of the parameter space
it is safe to neglect annihilations via the Higgs. However, if the DM mass is sufficiently close to the Higgs resonance,
these annihilations can become comparable to the annihilations within the dark sector. As such, we properly take into account all possible twin tau annihilations into the SM b-quark, gauge bosons, and Higgses. We take the annihilation rates 
from~\cite{Fedderke:2014wda} with the identification $\Lambda = \frac{f^2}{m_\htau}$. The thermally averaged annihilation rate is obtained from the 
cross sections via~\cite{Gondolo:1990dk},
\beq \label{eq:GondoloGelmini}
\langle \sigma v \rangle = \frac{1}{8m^4 T K_2^2(m/T)} \int_{4m^2}^\infty  \sigma(s) (s-4m^2) \sqrt{s} K_1(\sqrt{s}/T) ds~. 
\eeq

\begin{figure}
\centering 
\includegraphics[width=.49\textwidth]{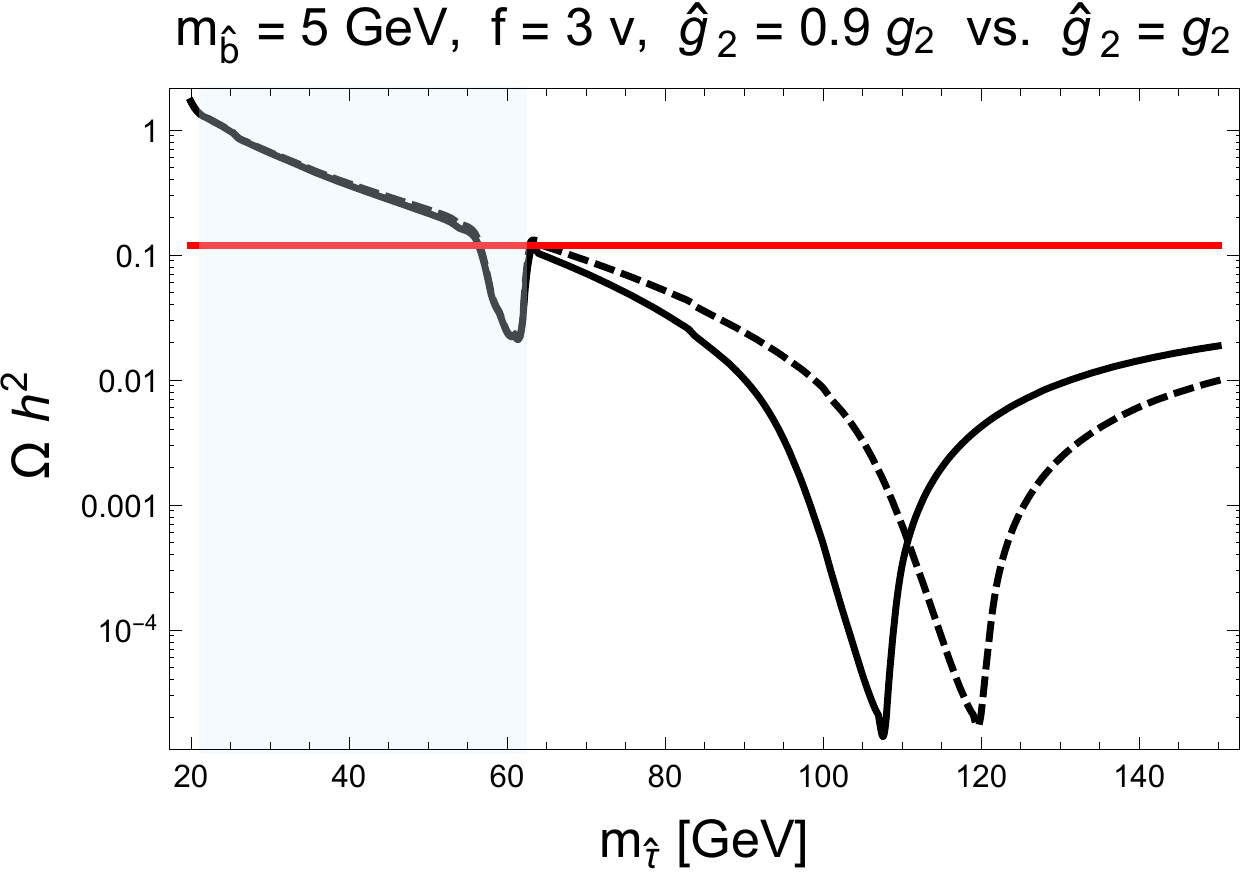}
\includegraphics[width=.49\textwidth]{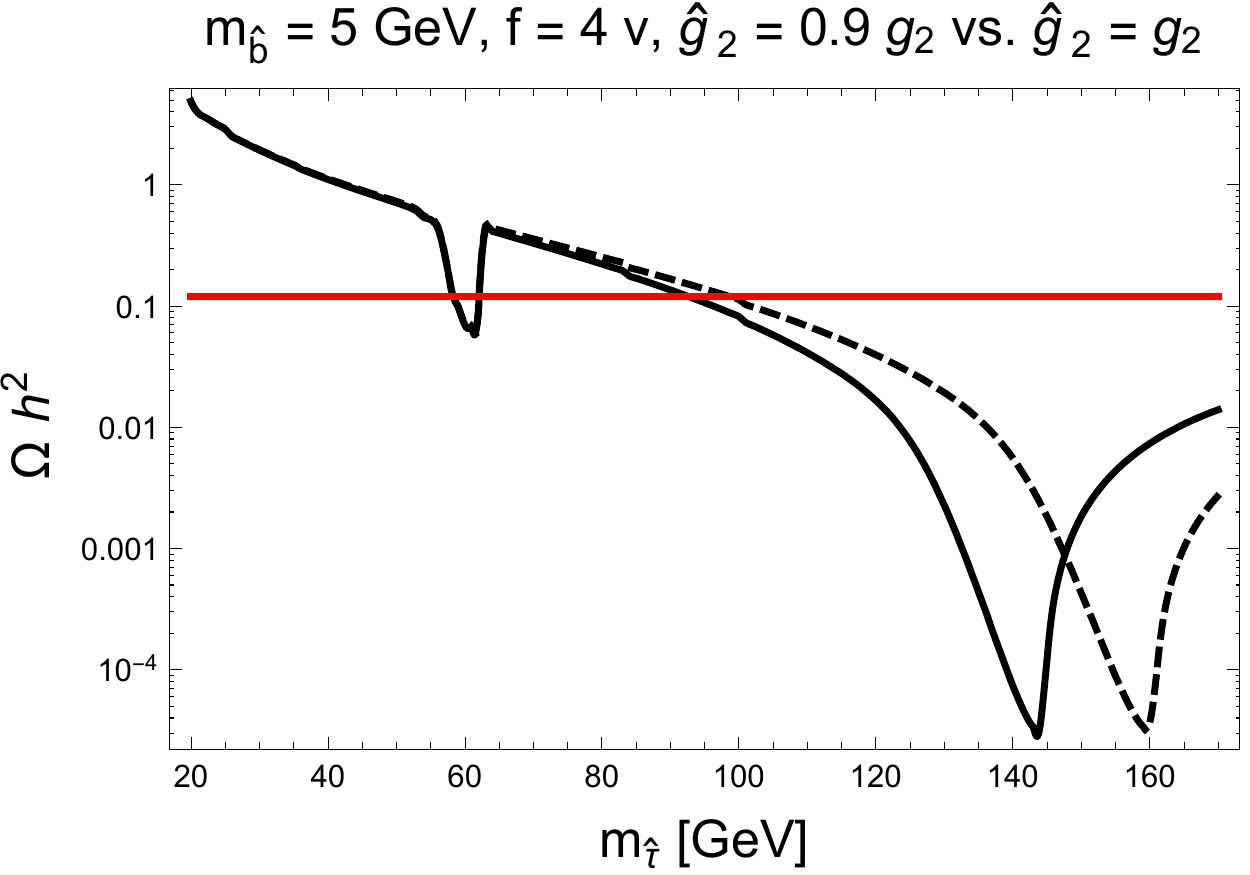}
\caption{Thermal relic abundance for different values of the twin $SU(2)$ gauge couplings. The dashed line stands for the twin gauge coupling equal to the visible one, and the 
solid line stands for smaller values of the $\hat g_2$, $\hat g_2 = 0.9 g_2$. The light blue shading indicates the region excluded by Higgs coupling measurements for $v/f = 1/3$ only. As we see, the mass at which we obtain the observed relic abundance is only mildly sensitive to this variation. }
\label{fig:gvariation}
\end{figure}

In general, the annihilation rate suggests that twin tau DM would be slightly overproduced for $m_\htau \approx m_\tau$. However, in the fraternal 
twin Higgs the masses of the twin bottoms and the twin taus are essentially free parameters. They need only satisfy the constraints that
\begin{enumerate}
\item The decay rate of the Higgs $h \to \hat{\tau}^+ \hat{\tau}^-$ should not be so large as to run afoul of bounds on the Higgs invisible width. For $v/f = 1/3$ this translates to a demand that the 
twin tau should be either lighter than $\sim$ 21~GeV or heavier than 62.5~GeV, whereas for smaller values of $v/f$ (i.e. $v/f \lesssim 1/4$) threshold suppression of $h \to \htau \htau$ leads to no irreducible bound on the mass of the twin tau.
\item The twin tau Yukawa should not be large enough to reintroduce the hierarchy problem at one loop via threshold corrections proportional to $\hat{y}_{\tau}$. This translates to a very mild demand that the twin tau should be lighter than
$\sim 200$~GeV. 
\end{enumerate}
Depending on the spectrum -- namely whether the annihilation into the twin b-quark is allowed or not -- the thermal relic abundance of the twin tau agrees with the observed dark matter abundance for twin tau masses between between 40 and 100~GeV. 
We show the expected 
thermal relic abundance as a function of mass in Fig.~\ref{fig:oh2noY}.  Clearly, the ``sweet spot'' mass is extremely sensitive to the precise value of $\frac{v}{f}$. As we have shown in Sec.~\ref{sec:FraternalTwin} values between, 21 and 62~GeV 
are disfavored by the constraints on invisible Higgs decays for $v/f = 1/3$, though allowed for smaller values of $v/f$. Most of the other thermal parameter space
is wide open and can vary from $\sim 62$~GeV all the way to more than 100~GeV. As we will see in the next section, almost all of this parameter 
space lies near (and below) the current LUX exclusion, providing tantalizing prospects for the next generation of direct detection experiments.  

Finally we notice that the variation of the $SU(2)$ twin gauge coupling does not change our results significantly. In the fraternal twin higgs scenario the twin 
$SU(2)$ coupling is allowed to vary within 10\% of the $SU(2)_L$ coupling. In the limit of zero-momentum cross sections we do not expect any dependence  
of the relic abundance on $\hat g_2$, but in principle it might become important in other regions of parameter space. It turns out that the region with the observed relic abundance is still fairly close to the zero-momentum approximation, and therefore the sensitivity to the exact value of $\hat g_2$ is minor. 
We illustrate this dependence in Fig.~\ref{fig:gvariation}.

\subsection{Gauging the twin hypercharge}
Thus far we have assumed that twin hypercharge is merely a global symmetry. However, it can also be gauged; this opens even more parameter space for light fraternal DM. 
We will further assume that that the twin photons are lighter than the 
DM candidate. They can be either massless, or twin electromagnetism can be broken
via the Stueckelberg mechanism at an arbitrarily high scale without reintroducing the 
hierarchy problem for new scalars acquiring small vevs.\footnote{It is also possible that the twin EM is broken via the Stueckelberg or Higgs mechanism at the a mass scale well above the DM mass. If near the scale of twin weak bosons, the twin photon opens up a new annihilation channel of comparable size. Significantly heavier twin photons are essentially irrelevant for dark matter. }

If the the hypercharge is gauged, the twin taus can also annihilate to 
the dark photons, on top of ``ordinary'' annihilation channels into twin neutrinos and twin bottoms. 
The differential cross section of the twin taus into a pair of dark photons is 
\beq \nonumber
\frac{d\sigma}{d\cos \theta} & = & \frac{2 \pi \hat \alpha^2}{\sqrt{s}} \frac{1}{\sqrt{s-4m_\htau^2}}
\Bigg( \frac{s + (s-4m_\htau^2) \cos^2 \theta}{4m_\htau^2 + (s-4m_\htau^2)\sin^2 \theta} + 
\frac{8 m_\htau^2}{4m_\htau^2 + (s-4m_\htau^2) \sin^2 \theta} - \\
&& - \frac{32 m_\htau^4}{(4m_\htau^2 + (s-4m_\htau^2)\sin^2\theta)^2}\Bigg)
\eeq 
where $\hat \alpha $ is the coupling of twin electromagnetism. In order to obtain 
a thermally averaged annihilation rate we numerically integrate this expression over 
$\cos \theta$ and further over $s$ according to~\eqref{eq:GondoloGelmini}.  If $m_\htau > m_\hb$, there is an additional annihilation channel into $\hb \bar \hb$ via the twin photon, but this channel
is also velocity suppressed.

If the annihilation into a pair of dark 
photons is the dominant annihilation mode, we would need $\hat \alpha \sim \cO(0.03 \times \alpha_{EM})$
in order to obtain the observed relic abundance for $m_{DM}\sim \cO(10~{\rm GeV})$~\cite{Hooper:2012cw}. Not surprisingly, this is also the order of magnitude of the twin electromagnetic coupling that would be necessary to open a parameter space for a light DM between 1~and 20~GeV. 
For much bigger couplings the dark matter is clearly underproduced in the entire allowed range of masses. On the other end, much smaller values of the twin electromagnetic coupling 
are also allowed. In this case the observed relic abundance can be accommodated primarily via the existing mechanisms considered earlier, or alternately via the twin photon for a very light (potentially below the GeV scale) dark matter candidate. The former possibility is identical 
to the twin sector without gauged hypercharge whatsoever. The latter is an interesting a logical possibility,
but we not discuss it in detail here owing to the challenges of  sub-GeV DM direct detection (see however~\cite{Essig:2012yx}).

\begin{figure}
\centering
\includegraphics[width=0.49\textwidth]{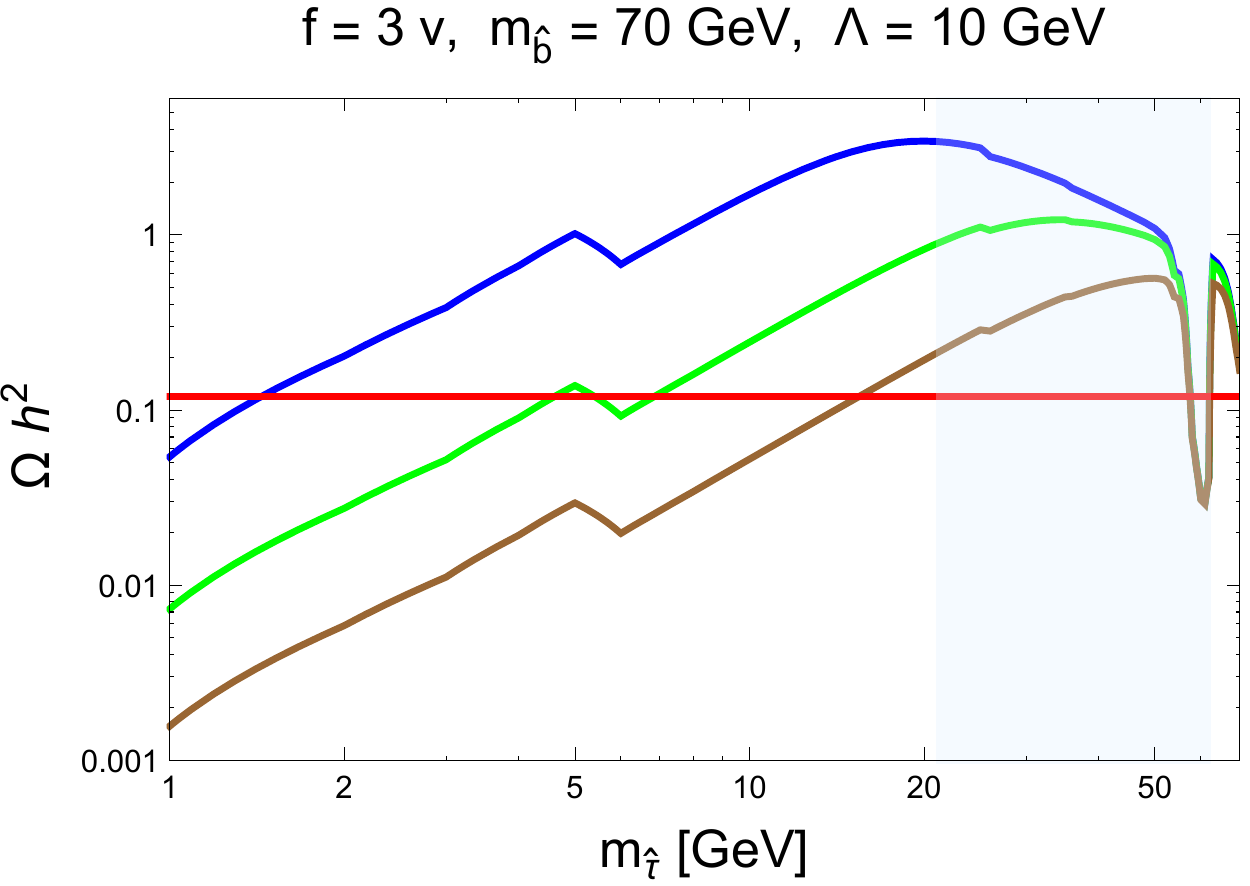}
\includegraphics[width=0.49\textwidth]{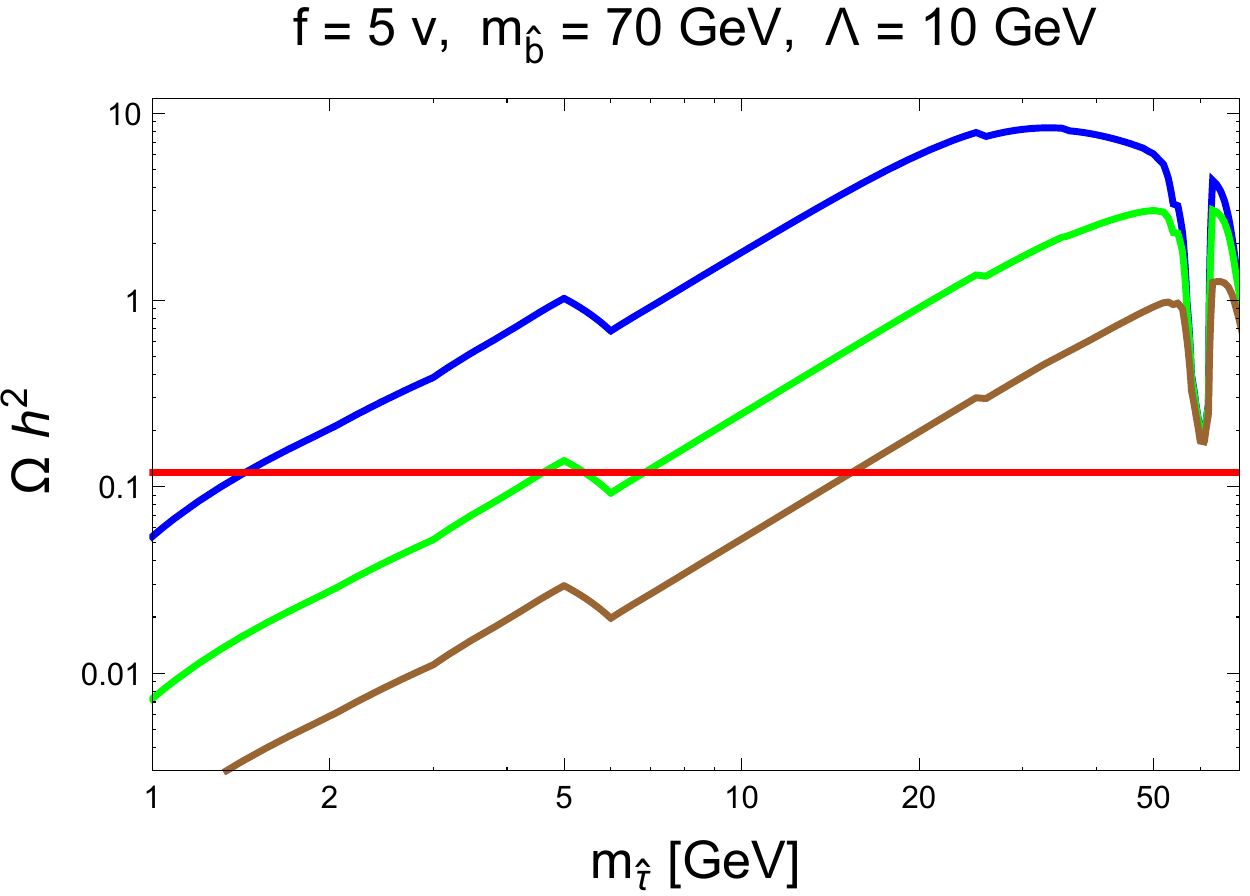}
\caption{Thermal relic abundance with gauged hypercharge. The blue, green and brown lines denote $\hat \alpha = 0.01\times ,\ 0.03\times ,\ 0.07\times \alpha_{EM}$ respectively. The red line indicates the measured relic 
abundance $\Omega h^2 =0.1187$. Left: Relic abundance for $v/f = 1/3$. The light blue region is excluded by Higgs coupling measurements. Right: Relic abundance for $v/f = 1/5$. Clearly, $v/f$ has a very minor effect on the low-mass region of the parameter space, which is dominated by annihilations
into dark photons.   The jump in relic abundance near the mass $\sim 5$~GeV is due to the fact that, for 
$m_\htau \lesssim 5$~GeV, decoupling happens after the (visible) QCD phase transition. }
\label{fig:relicdensityY}
\end{figure}

We show the predicted thermal relic abundance with gauged hypercharge in Fig.~\ref{fig:relicdensityY}. We see that twin electromagnetic
couplings of order $0.01 \times \alpha_{EM}$ open up the parameter space for all the DM masses  heavier than 1~GeV. Smaller gauge couplings will favor even lighter DM masses. Not surprisingly, the observed DM relic abundance is very hard to accommodate with $\hat \alpha \gtrsim 0.1 \times \alpha_{EM}$. 

Note that the existence of a light twin photon raises the prospect of maintaining thermal equilibrium between the Standard Model and twin sector until relatively late times, thereby altering the thermal history of the fraternal twin Higgs and increasing the contribution of the twin neutrino to $N_{\rm eff}$. 
However, the rate for interactions of the form $\hat \gamma \hat \gamma \leftrightarrow c \bar c$ mediated by the Higgs portal scales as 
\beq
\Gamma(\hat \gamma \hat \gamma \leftrightarrow c \bar c)  \sim y_c^2 \left( \frac{\hat \alpha}{6\pi}\right)^2 \left( \frac{v}{f}\right)^2 \left( \frac{T}{f}\right)^2
\left( \frac{T}{m_h}\right)^4 T~.
\eeq  
For the relevant values of $\hat \alpha \sim 10^{-2} \alpha_{EM}$ and temperatures $T \sim 1$~GeV this is subleading relative to the contributions discussed in Sec.~\ref{sec:ThermalHistory}. Kinetic mixing also potentially connects the Standard Model and twin sectors, but this contribution is negligible for the range of kinetic mixing parameters compatible with direct detection, as we will see in Sec.~\ref{sec:detection}.

\subsection{$\z_n$ generalizations }

All the ideas described in the previous subsection can be easily realized in generic $\z_n $ theories as  described in Sec.~\ref{sec:FraternalTwin}. Of course, in a generic $\z_n$ model we have potentially coexistent DM in all 
$(n-1)$ different sectors. {\it A priori} the different sectors may even have different kinds of DM, for example asymmetric DM in one 
sector and thermal dark matter in another sector. Exploring the full set of possibilities is beyond the scope of this study; here we will concentrate on the cleanest possibility, namely thermal twin tau DM coexisting in the different twin sectors. 

In order to illustrate this point, we will concentrate on a simplest  possible extension of the fraternal twin Higgs, the $\z_3$ model. 
All other extensions closely follow the same logic. To illustrate the main point we will make one more simplification and assume that the VEVs of the Higgses in the non-SM sectors are equal, namely $v_B = v_C \approx \frac{f}{\sqrt{2}}$. Note that an even broader range of possibilities arise when this assumption is relaxed.

In the absence of gauged twin hypercharges, the annihilation cross section for the DM in each different sector is 
\beq
[\sigma v]_{\htau \htau\to \hnu \hnu} \approx \frac{1}{4\pi} \left( \frac{v}{v_i}\right)^4 G_F m_\htau^2~. 
\eeq
since masses of the weak gauge bosons in each sector are $\sim g v_i / 2$ rather than $\sim g f / 2$.
There is of course a potential extra contribution from annihilation into $\hb \bar \hb$ if the twin b-quarks are lighter than the twin taus in the dark sector,  analogous to Eq.~\ref{eq:AnnBs} up to the appropriate change $f \to v_i$. 

As we keep the scale of spontaneous $SU(2n)$ breaking $f$ constant and increase the number of sectors $n$, we \emph{increase} the 
annihilation cross sections, such that these generalizations  of the twin Higgs will in general favor slightly lighter DM
that the original fraternal twin Higgs model.    

\begin{figure}[t]
\centering
\includegraphics[width=.49\textwidth]{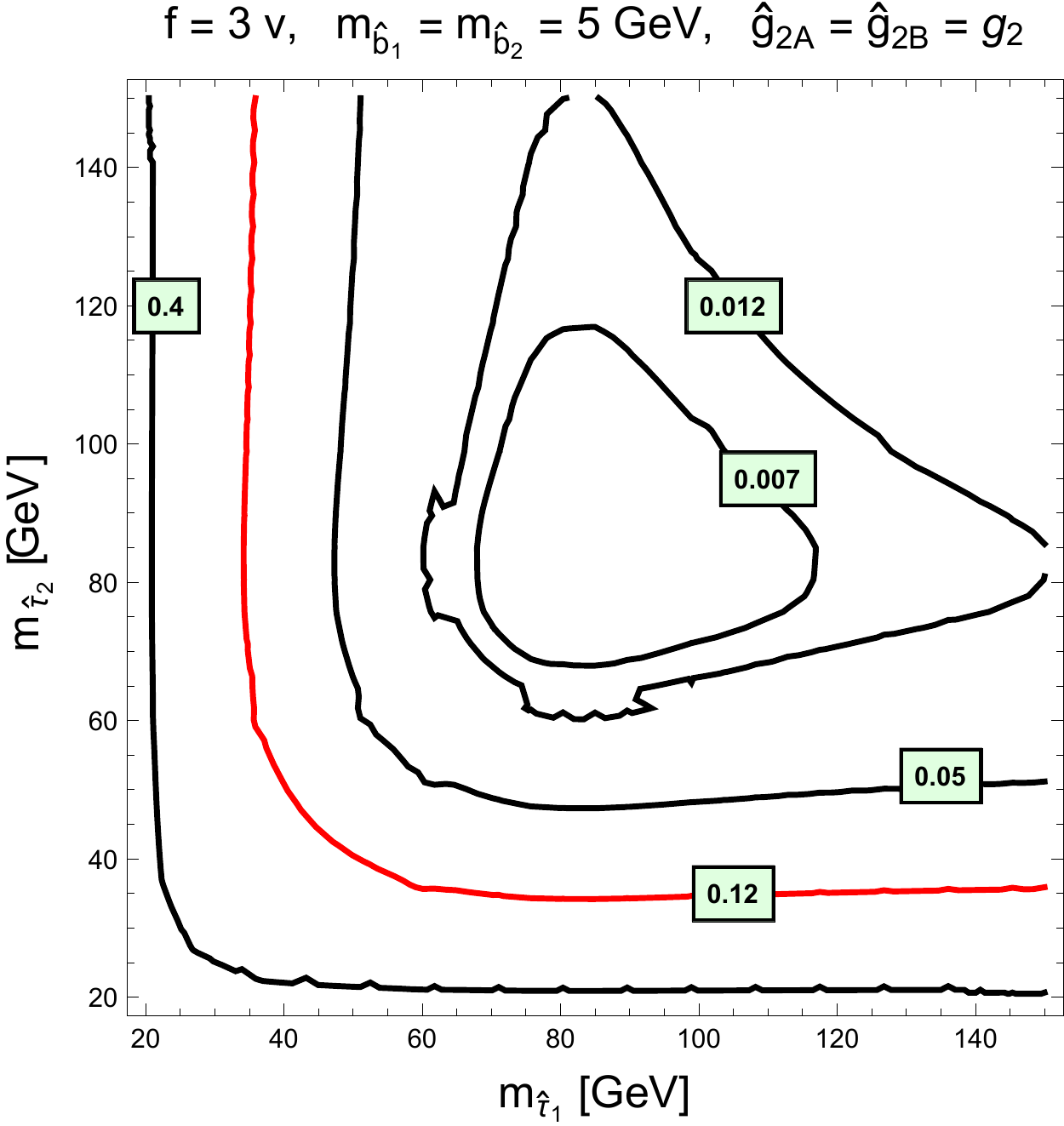}
\includegraphics[width=.49\textwidth]{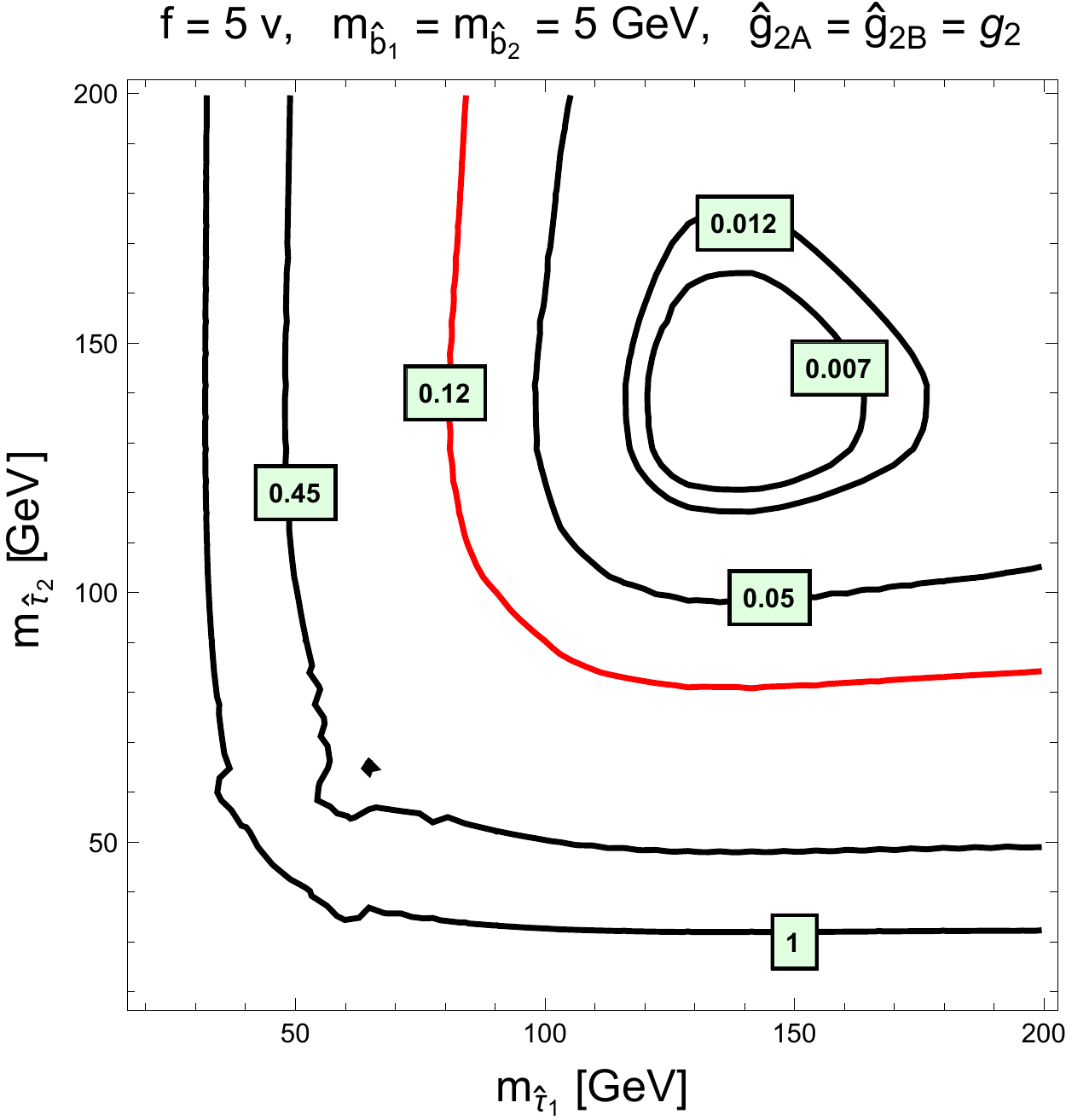}
\caption{Thermal relic abundance in a $\z_3$ model as a function of twin tau masses. We assume 
$v/f=1/3$ (left-hand side) and $v/f = 1/5$ (right-hand side) and equal scales of breaking for the two twin $SU(2)$s in both cases. 
The thick black contours indicate the predicted 
thermal relic abundance, while the measured relic abundance $ \Omega h^2 = 0.1187$ is shown by a red contour. 
 }
\label{fig:z3Relic}
\end{figure}

Of course the masses of the twin taus in the different sectors are {\it a priori} independent and uncorrelated. To accommodate this, we allow different masses for the twin taus in different sectors. The expected total dark matter relic abundance
and relative abundance of the twin taus is shown in Fig.~\ref{fig:z3Relic}, assuming
relatively light twin bottoms and twin QCD scale (as we have seen, the dependence on these parameters is very mild). 
Comparing to Fig.~\ref{fig:oh2noY} we see that in general the $\z_n $ generalization of the fraternal twin Higgs prefers lighter DM values. 

As we have already mentioned, in the simplest $\z_2$ scenario dark matter masses between  $21-62.5$ ~GeV are in tension with constraints on the non-SM Higgs width for $v/f = 1/3$, though these bounds are ameliorated by smaller $v/f$. Here, a highly simplified  $\z_3$ scenario ($v_2 = v_3 \gg v_{SM}$) predicts that the dominant DM component is $\sim$40~GeV in very large portions of the parameter space. This is still in tension with Higgs coupling fits, though $\z_n$ generalizations with $n > 3$ are likely to provide light DM candidates compatible with coupling fits. Unsurprisingly, the dominant relic is the lightest twin tau, while the heavier twin tau usually constitutes a small or even negligible component of the dark matter.

In this discussion we have assumed that twin hypercharge is not gauged in either hidden sector. It can also be gauged, in which case our previous discussion of gauged twin hypercharge generalizes accordingly, allowing even lower masses for the thermal DM candidates.

\section{Comments on Direct Detection}  
\label{sec:detection}
Although the thermal DM candidate predominantly annihilates via the twin weak force, these couplings do not translate into couplings with the visible sector. Rather, the leading coupling between the DM and the visible sector arises upon integrating out the heavy twin Higgs mode, as given in Eq.~\ref{eq:yukawa}. As such, direct detection signals are entirely dominated by the Higgs portal. The coupling to SM baryonic matter via the Higgs portal is spin-independent and velocity-independent. 

The implications of the fermionic Higgs portal for direct detection were analyzed in detail in Ref.~\cite{LopezHonorez:2012kv,Fedderke:2014wda}, and we will largely follow the discussion therein.  The scattering cross section of the DM on the proton is 
\beq\label{eq:DD}
\sigma(\htau p \to \htau p) = \frac{1}{\pi} \left( \frac{m_\htau v}{f^2} \right)^2 g_{Hp}^2 \frac{m_{red}^2}{m_H^4}~,
\eeq
where $m_{red}$ is reduced mass of the proton and the DM. 
The coefficient $g_{Hp} $ is given by
\beq
g_{Hp} = \frac{m_p}{v}\left( \sum_{q} f_q^{(p)} + \frac{2}{9}\left( 1 - \sum_{q}f_q^{(p)}\right) \right)~. 
\eeq
The coefficients $f_{q}$ can be found in~\cite{Belanger:2008sj}, for more recent estimates see~\cite{Crivellin:2013ipa}. 

Evaluated numerically, the expected cross sections are just slightly below $10^{-45}$~cm$^2$. This places direct detection cross sections right on the edge of the current LUX exclusion bounds~\cite{Akerib:2013tjd}. Of course, the precise value of the scattering cross section is very sensitive to both the $v/f$ parameter and the DM mass, inviting more careful study. 

\begin{figure}
\centering
\includegraphics[width=.7\textwidth]{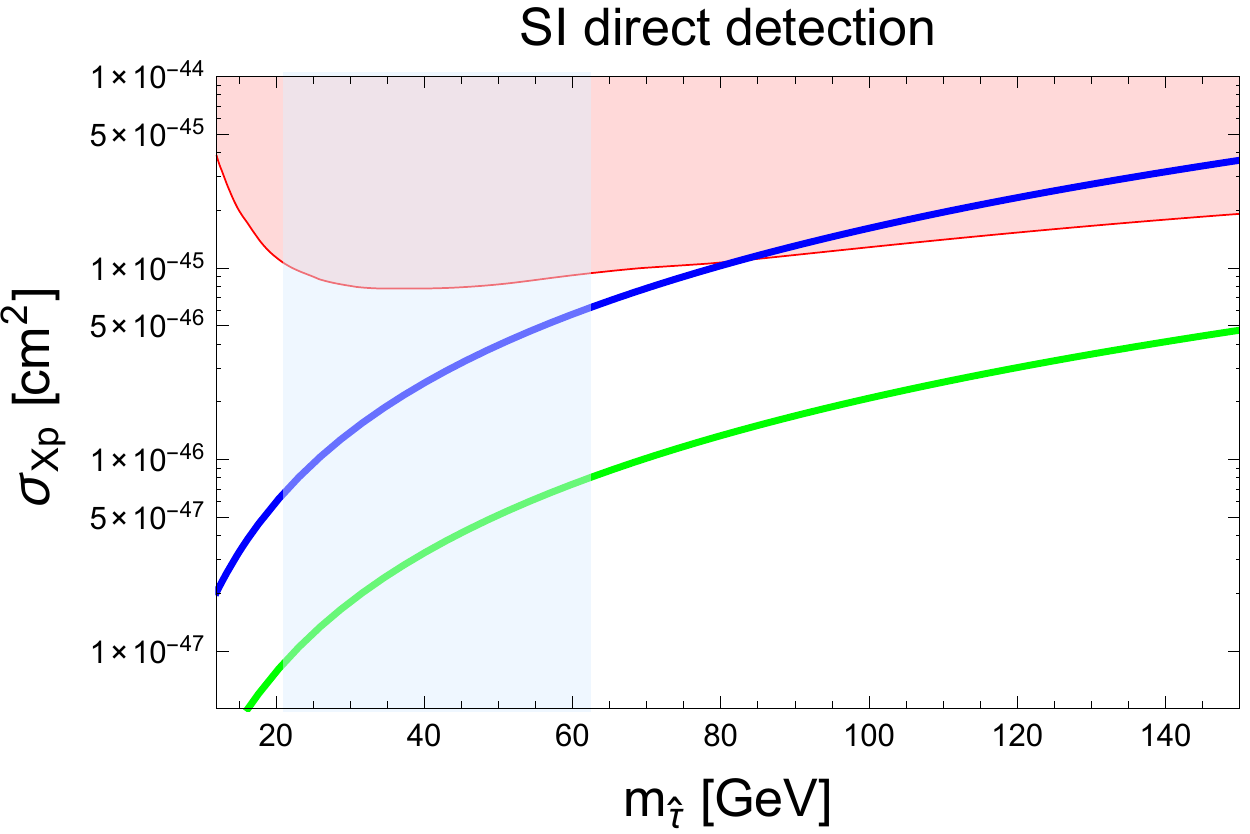}
\caption{Predicted nuclear scattering cross sections of the twin tau DM on protons. The green 
line corresponds to $\frac{v}{f} = \frac{1}{5}$ and 
the blue line corresponds to $\frac{v}{f} = \frac{1}{3}$. The red region is excluded by the current  LUX  bound~\cite{Akerib:2013tjd}. The light blue region is excluded by Higgs coupling measurements for $v/f = 1/3$ only.} 
\label{fig:LUX}
\end{figure}

We present our detailed numerical results in Fig.~\ref{fig:LUX}. While for $\frac{v}{f} = \frac13$ some portion of parameter space is already excluded by LUX results,
the bounds are significantly attenuated for larger values of $v/f$. Essentially the entire parameter space is allowed for 
$\frac{v}{f}=\frac15$. However, the relevant cross sections are not far from the region currently explored by LUX; most or all of the relevant parameter space will be decisively explored by the next generation of direct detection experiments. This strongly motivates further direct detection experiments, which may be considered as complimentary to collider probes of neutral naturalness. 

As we consider theories with gauged hypercharge, we of course open up more parameter space at the cost of reduced predictivity. By gauging the hypercharge we allow much lighter DM masses relative to the minimal scenario. Such light DM is very difficult for direct detection experiments due to tiny recoil energies. However, the interaction cross sections can be significantly enhanced compared to the minimal fraternal twin Higgs scenario. On top 
of the Higgs portal, an additional lower-dimensional portal opens up, namely kinetic mixing:
\beq
\cL \supset \frac{\kappa}{2} F^{\mu \nu} {F'}_{\mu\nu}
\eeq
Although we know that in the low-energy fraternal effective theory the kinetic mixing parameter $\kappa$ is stable to radiative corrections up three loops, it may be generated by unspecified physics in the UV completion and should be treated as a free parameter. This kinetic mixing yields an extra contribution to the spin-independent scattering between DM and protons that depends strongly on the mass of the dark photon. For dark photon masses above 10~MeV, this contribution is~\cite{Hooper:2012cw}
\beq
\sigma_{\htau p \to \htau p} = \frac{g_2^2 \sin \theta_W^2 \hat g_{EM}^2 \kappa^2 m_\htau^2 m_p^2}{\pi
m_{\hat \gamma}^4 (m_\htau+m_p)^2}
\eeq
Around $\sim$10~MeV the mass of the dark matter becomes comparable to the typical momentum exchange between the proton and the DM, and such that more careful treatment of the momentum dependence is required (see~\cite{Fornengo:2011sz} for detailed discussion). The contribution to direct detection experiments is saturated for $m_{\hat \gamma} \approx 10$~MeV, such that our results for 
the mediator mass of 10~MeV will also be for lighter dark photon masses.

\begin{figure}
\centering
\includegraphics[width=0.49\textwidth]{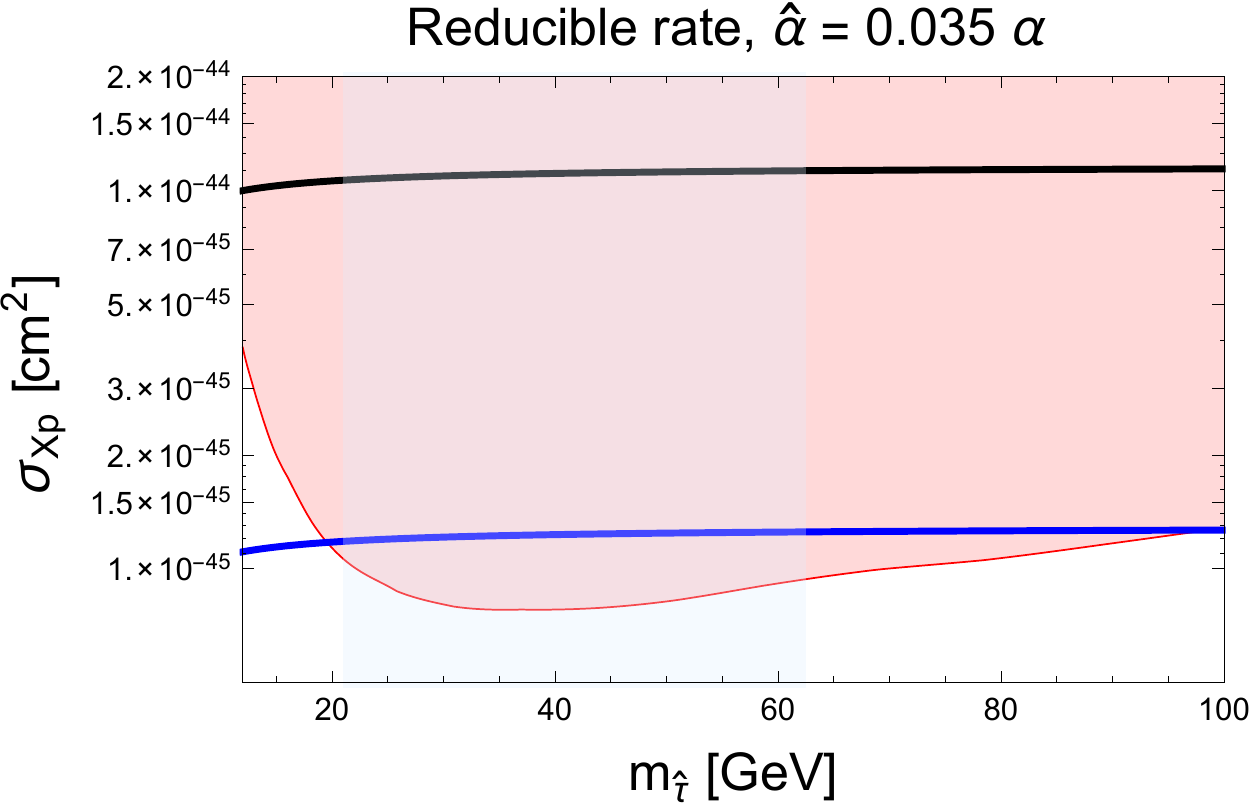}
\includegraphics[width=0.47\textwidth]{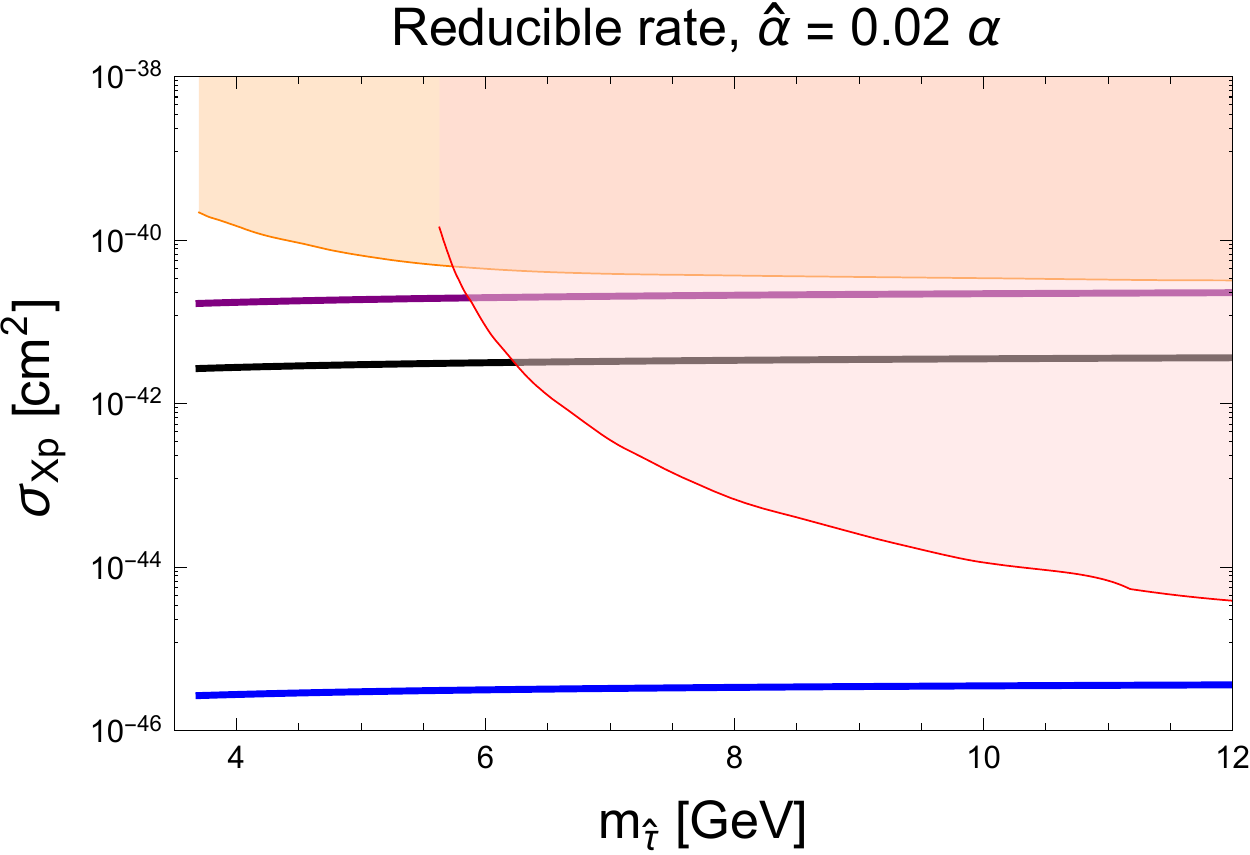}
\caption{Reducible twin tau - proton scattering cross sections in the presence of gauged twin hypercharge. Left: The expected nuclear scattering cross sections for ``heavy'' (20-100 GeV) $\htau$ DM with dark photon mass $m_{\hat \gamma} = 100$~MeV. The blue line indicates the scattering cross section for kinetic mixing parameter $\kappa = 10^{-8}$, while the black line indicates the scattering cross section for $\kappa = 4 \times 10^{-8}$. Current LUX bounds are in red. The light blue region is excluded by Higgs coupling measurements for $v/f = 1/3$ only. Right: The expected nuclear scattering cross sections for ``light'' (4-12 GeV) $\htau$ DM, suitable for thermal abundance in the presence of gauged hypercharge. The blue line corresponds to the scattering cross section for 
 $m_{\hat \gamma} = 100$~MeV, $\kappa = 10^{-8}$; the black line to $m_{\hat \gamma} = 100$~MeV, 
$\kappa = 10^{-6}$;  and the purple line to $m_{\hat \gamma} \leq 10$~MeV, $\kappa = 2.5\times 10^{-8}$. The red region is excluded by 
LUX, while the orange region indicates exclusion due to CDMS bounds~\cite{Agnese:2013jaa}.}
\label{fig:ReducibleRates}
\end{figure}

We show the reducible\footnote{``Reducible'' in the sense of ``unrelated to naturalness in the fraternal twin Higgs, but potentially present''.} rates for various values of the dark photon and kinetic mixing mass in Fig.~\ref{fig:ReducibleRates}. Note that for DM heavier than about 10~GeV, the kinetic 
mixing should be below $\sim10^{-8}$ in order to avoid direct detection limits; this also ensures that the kinetic mixing portal is too small to keep the twin sector in equilibrium with the SM below the QCD phase transition. Demanding such a small
kinetic mixing might entail some degree of fine-tuning but depends on the details of the UV theory. 

However, as we have seen, in the presence of gauged twin hypercharge the correct DM relic abundance is typically obtained for DM masses below 10~GeV where the 
direct detection constraints are much weaker.  For a 100~MeV dark photon, we can allow kinetic mixing on the order of $\kappa \sim 10^{-6} - 10^{-8}$ 
depending on the exact mass, and even dark photons which saturate the cross section (with masses around or below 10~MeV) can still have kinetic mixings above $10^{-8}$. At least within the low-energy fraternal twin Higgs scenario this does not demand any excessive fine-tuning of the kinetic mixing parameter. Most of the kinetic 
mixings which are not fine-tuned to be overly small are clearly within two orders of magnitude of the best 
available bounds (from LUX above 6~GeV and SuperCDMS~\cite{Agnese:2013jaa} 
for the lighter DM masses). This clearly
motivates further exploration of the light DM parameter space.  

Finally we comment on the $\z_n$ extensions of the fraternal twin Higgs. Because these scenarios entail coexisting dark matter species and each sector contributes some fraction of the total relic density (generically with different masses and interaction rates), a proper 
reinterpretation of LUX results would be required to determine whether a given scenario is excluded or not. Such a reinterpretation is beyond the scope of this paper, but we can clearly see that direct detection bounds are attentuated in even in the minimal $\z_3$ scenario, accommodating thermal dark matter consistent with LUX bounds. If the DM mass differs significantly between various sectors, 
the heaviest components of the dark matter usually constitute a small fraction of the DM, yet deposit more energy in the detector. This leads to an amusing possibility that may arise in next-generation of direct detection experiments, wherein multiple dark matter populations give rise to diverse recoil spectra.

\section{Conclusions and Outlook} 
\label{sec:conclusions}

The WIMP miracle provides a suggestive mechanism for generating the observed dark matter relic abundance that dovetails nicely with new stable states predicted by conventional approaches to solving the hierarchy problem. However, the onward march of null results in both dark matter direct detection experiments and LHC searches for new physics has placed these intertwined paradigms under some degree of tension.  

Reconciling LHC null results with a natural weak scale has recently led to exploration of ``neutral naturalness'' -- theories in which the (little) hierarchy problem is addressed by new degrees of freedom that are partially or entirely neutral under the Standard Model.  These include the mirror twin Higgs, fraternal twin Higgs, and generalizations thereof. Such models are merely examples of a more general mechanism, in which additional hidden sectors related by discrete symmetries improve the naturalness of the weak scale through Higgs portal-type interactions. In all such theories, the mass scales of the hidden sector(s) must be parametrically close to the weak scale, while the couplings must be closely related to their Standard Model counterparts. Much as conventional approaches to the hierarchy problem have lent themselves to conventional WIMP dark matter candidates, this suggests a ready generalization of the WIMP miracle in theories of neutral naturalness.

In this paper we have demonstrated that these approaches to the hierarchy problem can also generically give rise to a ``fraternal WIMP miracle'', in which SM-neutral states in the twin sector(s) are stabilized by gauge or global symmetries. Although the relevant scales for freeze-out and direct detection are not directly set by the weak scale, they are constrained to lie near the weak scale by naturalness considerations. The thermal relic abundance of fraternal dark matter candidates is largely set by annihilation into twin neutrinos via massive twin weak bosons, with an annihilation cross section parametrically related to the weak scale by powers of $f/v \sim {\rm few}$. Their direct detection cross sections are instead set by Higgs portal interactions with parametrically distinct scaling, which generically puts them in a regime not currently probed by direct detection but accessible to future experiments. 

We have analyzed in detail the thermal relic abundance of dark matter candidates in fraternal twin Higgs models and their generalizations, finding that the twin tau makes an ideal dark matter candidate with a viable parameter space consistent with existing bounds on fraternal scenarios. We have further studied the parameter space for direct detection, finding that much of the parameter space for fraternal dark matter with thermal relic abundance is compatible with existing direct detection bounds. 

In the minimal fraternal twin Higgs within the range $v/f = 1/3-1/5$ favored by current bounds and naturalness considerations, the twin tau obtains the correct thermal abundance consistent with Higgs coupling limits for $m_{\hat \tau} \sim 63 - 150$ GeV, assuming twin hypercharge is not gauged. A somewhat wider parameter space is available in generalizations of the minimal model with multiple sectors. If twin hypercharge is gauged the thermal abundance depends sensitively on the twin hypercharge gauge coupling, with suitable thermal candidates in the range of $m_{\hat \tau} \sim 1-20$ GeV for twin electromagnetic couplings of order $0.01 \; \times \; \alpha_{EM}$. In the absence of twin hypercharge, the direct detection cross sections are within an order of magnitude of current LUX limits for $v/f = 1/3-1/5$, with 63 GeV $\lesssim m_{\hat \tau} \lesssim 80$ GeV allowed for $v/f = 1/3$ and all relevant masses allowed for $v/f = 1/5$. If twin hypercharge is gauged, the low-mass region consistent with thermal abundance is likewise within several orders of magnitude of current limits but highly sensitive to kinetic mixing and the twin hypercharge gauge boson mass.

There are a variety of interesting future directions. We have focused on the simplest scenarios involving the minimal fraternal Twin Higgs and simple generalizations to multiple twin sectors. However, the fraternal WIMP miracle is a generic feature of hidden sectors that protect the weak scale through approximate discrete symmetries, with a range of parametric freedom for the dark matter candidate(s) that merits further investigation. In this work we have also focused on gauged or global twin hypercharge as a stabilization symmetry for dark matter. In principle, conserved baryon and lepton numbers in the twin sector may also serve as stabilization symmetries, giving rise to asymmetric dark matter candidates. A suggestive possibility would be to simultaneously seed Standard Model and twin sector baryon asymmetries, potentially explaining the observed near-coincidence of dark matter and baryon densities. More broadly, the paradigm of neutral naturalness suggests diverse possibilities for new hidden sectors with scales parametrically close to the weak scale, opening the door to new  connections between dark matter and the hierarchy problem.

\acknowledgments
We thank Marco Farina, Roni Harnik, Simon Knapen, Pietro Longhi, John March-Russell, Matthew McCullough, and Raman Sundrum for useful conversations, and Isabel Garcia Garcia, Robert Lasenby, and John March-Russell for helpful comments on the manuscript. We particularly thank Anupam Mazumdar and Matt Strassler for collaboration and insight in the early stages of this work. NC is supported by the Department of Energy under the grant DE-SC0014129.

\vspace{0.5cm}
\noindent {\bf Note added:} While this paper was in preparation we learned of related work by Garcia Garcia, Lasenby, and March-Russell \cite{JMR1,JMR2} and by Farina \cite{Farina}. We are grateful to these authors for correspondence and coordination in publication.

\bibliography{fraternal}
\bibliographystyle{jhep}
\end{document}